# On Distributed Linear Estimation With Observation Model Uncertainties

Alireza Sani, *Student Member, IEEE,* Azadeh Vosoughi, *Senior Member, IEEE,*

*Abstract*—We consider distributed estimation of a Gaussian source in a heterogenous bandwidth constrained sensor network, where the source is corrupted by independent multiplicative and additive observation noises, with incomplete statistical knowledge of the multiplicative noise. For multi-bit quantizers, we derive the closed-form mean-square-error (MSE) expression for the linear minimum MSE (LMMSE) estimator at the FC. For both error-free and erroneous communication channels, we propose several rate allocation methods named as *longest root to leaf path*, *greedy* and *integer relaxation* to ($i$) minimize the MSE given a network bandwidth constraint, and ($ii$) minimize the required network bandwidth given a target MSE. We also derive the Bayesian Cramér-Rao lower bound (CRLB) and compare the MSE performance of our proposed methods against the CRLB. Our results corroborate that, for low power multiplicative observation noises and adequate network bandwidth, the gaps between the MSE of our proposed methods and the CRLB are negligible, while the performance of other methods like *individual rate allocation* and *uniform* is not satisfactory.

## I. INTRODUCTION

Large-scale sensor networks, consisting of battery operated devices with limited sensing, computation, and communication capabilities, can impact various applications, including environmental monitoring, surveillance, target tracking, and can be utilized to perform challenging tasks such as distributed estimation (DES). For DES, sensors send their locally processed observations to a Fusion Center (FC), that is tasked with estimating an unknown parameter, via fusing the collective information received from the sensors, such that the estimation error at the FC is minimized, subject to practical network resource constraint(s), such as transmit bit or transmit power constraint. In this work, we address two important problems pertaining bandwidth-constrained DES in a sensor network: ($i$) given a network bandwidth constraint, we investigate the quantization rate allocation schemes that minimize Mean Square Error (MSE) at the FC, ($ii$) given a target MSE at the FC, we explore the quantization rate allocation schemes that minimize the required network bandwidth.

DES has been of great research interest in signal processing society [2]–[20]. Several researchers have studied quantization design, assuming that sensors' quantized observations are sent over bandwidth-constrained (otherwise error-free) communication channels [2], [3], [6]–[8]. Note that placing a cap on the total bandwidth can further improve energy efficiency [21]–[25], because data communication is a major contributor to the network energy consumption. In particular, [2] designed

Part of this research was presented at the IEEE 82nd Vehicular Technology Conference, 2015 [1]. This research is supported by the NSF under grants CCF-1341966 and CCF-1319770.

quantizers that maximize Bayesian Fisher information (FI). [3] designed identical one-bit quantizers that minimize Cramér-Rao lower bound (CRLB). For a homogeneous network with Maximum-Likelihood Estimator (MLE) at the FC, [7] investigated one-bit quantizers. For estimating a random parameter with uniform variable rate quantizers, [8] investigated the tradeoff between fine quantization of observations of only a few sensors and coarse quantization of observations of as many sensors as possible, as well as its effect on FI, subject to a network bandwidth constraint. For a heterogeneous network with best linear unbiased estimator (BLUE) at the FC, [6] proposed a suboptimal bit allocation scheme that minimizes MSE, subject to a network bandwidth constraint. For a homogeneous network with MLE at the FC, [5] designed one-bit quantizers, where each sensor optimizes its quantizer threshold. For DES of an unknown deterministic parameter vector, [4] proposed a joint compression-quantization scheme that first reduces a sensor's observation to a scalar and then applies a one-bit quantizer, such that CRLB is minimized. For a homogeneous network with one bit quantizers at the sensors, [10] designed a universal linear estimator at the FC, assuming that the deterministic unknown and the additive observation noise are in symmetric bounded intervals around zero.

An inherently related problem to DES is the quadratic Gaussian CEO problem in information theory [26]–[29], where a team of agents observe a Gaussian source, corrupted by independent additive Gaussian noises. Agents communicate their coded messages over error-free channels to the CEO (FC), that is tasked with estimating the Gaussian source with minimal MSE distortion. For this problem [26] conjectured a rate-distortion region, where the decay rate of MSE is inversely proportional to total source coding rate of agents when the number of agents goes to infinity. Later, [28] found the rate-distortion region explicitly for arbitrary number of agents. It is worthwhile to emphasize the differences between DES and the quadratic Gaussian CEO problem. Addressing the CEO problem (from the information theoretic perspective) is built upon fundamental assumptions that are often challenging to be satisfied in practical settings and hence the produced results (including the rate allocation schemes in [27], [29]) cannot be directly applied to sensor networks. In particular, sensors sample their observations continuously and use infinite-length complex source codes (jointly typical sequences) to compress their samples into bits, and the FC utilizes the best decoder (estimator), in terms of minimizing MSE. These assumptions disregard several key characteristics of sensor networks, including hardware complexity, computational capability, power, and delay constraints. In contrast, DES keeps the delay and computational complexity low, as each sensor has one sample (of an unknown or a vector of unknowns) to quantize.

The bulk of literature on DES assumes that the observation (sensing) model is completely specified [2]–[17], while some assume that the statistical knowledge of the additive observation noise is incomplete (noise model uncertainty) [7], [10]. Centralized estimation with multiplicative and additive observation noises has been studied before in [30]–[32]. Despite its great importance, few researchers have studied DES with both multiplicative and additive observation noises. DES with Gaussian multiplicative and additive observation noises, one-bit quantizers at the sensors and MLE at the FC has been investigated very recently in [33], [34], respectively, for vector and scalar unknown deterministic parameters. We also note that most DES literature is focused on one-bit quantization [2]–[5], [7], [9], [10], [15], [33], [34], assuming that the dynamic range of the unknown is equal to or less than that of the additive observation noise [10]. Interestingly, [7] argued that there can be a significant gap between the CRLB performance based on one-bit quantization and the clairvoyant benchmark (unquantized observations are available at the FC), when the dynamic range of the unknown is large with respect to the additive observation noise variance. Also, [33] reported that in presence of multiplicative noise, low power additive observation noise can negatively impact the performance of MLE based on one-bit quantization. The works in [7], [33] motivate us to consider DES based on multi-bit quantization.

**Our Contributions**: We consider DES of a Gaussian source, corrupted by independent multiplicative and additive observation noises, in a heterogeneous bandwidth-constrained network. Similar to [6], [7], we choose the total number of quantization bits as the measure of network bandwidth. Different from [33], [34], we assume that the distribution of the multiplicative observation noise is unknown and only its mean and variance are known (observation model uncertainties). To overcome the limitations caused by the observation model uncertainties, the FC employs linear minimum MSE (LMMSE) estimator to fuse the quantization bits received from sensors over orthogonal channels. We consider both error-free and erroneousness communication channels [21], [35], using binary symmetric channel (BSC) model. We focus on the effects of multiplicative noise, observation model uncertainties and quantization errors on the accuracy of estimating the Gaussian source. We derive a closed-form expression for the MSE of LMMSE estimator and consider two system-level constrained optimization problems with respect to the sensors' quantization rates: in (**P1**) we minimize the MSE given a network bandwidth constraint, and in (**P2**) we minimize the required network bandwidth given a target MSE. To address these two problems, we propose several rate allocation schemes. In addition, we derive the Bayesian CRLB and compare the MSE performance of the proposed schemes against the CRLB.

## II. SYSTEM MODEL AND PROBLEM STATEMENT

We consider a network with $K$ spatially distributed heterogeneous sensors and a FC, where the FC is tasked with estimating a realization of a Gaussian source $\theta \sim \mathcal{N}(0, \sigma_\theta^2)$, via fusing the collective received data from all sensors. Each sensor makes a noisy observation of $\theta$, where *both multiplicative and additive observation noises* are involved. Let $x_k$ denote the scalar noisy observation of $\theta$ at sensor $k$. We assume the following observation model:

$$x_k = h_k \theta + n_k, \quad \text{for} \quad k = 1, ..., K, \quad (1)$$

where $h_k$ and $n_k$ are multiplicative and additive observation noises, respectively. Also, $h_k, n_k, \theta$ are uncorrelated. We assume $n_k \sim \mathcal{N}(0, \sigma_{n_k}^2)$, $\mathbb{E}\{h_k\} = 1 \ \forall k$, and[1] $\text{var}(h_k) = \sigma_{h_k}^2$. Sensor $k$ employs a uniform quantizer with $M_k$ quantization levels and quantization step size $\Delta_k$. The quantizer maps $x_k$ into a quantization level $m_k \in \{m_{k,1}, ..., m_{k,M_k}\}$, where $m_{k,i} = \frac{(2i-1-M_k)\Delta_k}{2}$ for $i = 1, ..., M_k$. We assume $x_k$ lies in the interval $[-\tau_k, \tau_k]$ almost surely, for some reasonably large value of $\tau_k$, and we let $\Delta_k = \frac{2\tau_k}{M_k - 1}$. These imply that the uniform quantization mapping rule can be described as the following: if $x_k \in [m_{k,i} - \frac{\Delta_k}{2}, m_{k,i} + \frac{\Delta_k}{2}]$, then $m_k = m_{k,i}$, if $x_k \geq \tau_k$, then $m_k = \tau_k$, and if $x_k \leq -\tau_k$, then $m_k = -\tau_k$. Following quantization, sensor $k$ maps the index $i$ of $m_{k,i}$ into a bit sequence of length $r_k = \log_2 M_k$ and sends $r_k$ bits to the FC. Sensors transmit over orthogonal bandwidth-constrained error-free communication channels. Error-free communication channel model has been used before in several classical works on DES, examples are [1]–[8]. In Section VIII we extend our analytical results to the case where these channels are modeled as independent BSC with different error probabilities. To capture the network bandwidth constraint we assume $\sum_{k=1}^{K} r_k \leq B_{tot}$. In the absence of knowledge of joint distribution of $m_k$'s and $\theta$, we resort to the linear minimum MSE (LMMSE) estimator [36] to form the estimate $\hat{\theta} = \boldsymbol{Gm}$ at the FC, where $\boldsymbol{G}$ is the $1 \times K$ linear estimation operator and $\boldsymbol{m} = [m_1, ..., m_K]^T$ is the vector of transmitted quantization levels for all sensors. The LMMSE estimator has a low computational complexity and only requires the knowledge of moments $\mathbb{E}\{\theta \boldsymbol{m}^T\}$ and $\mathbb{E}\{\boldsymbol{m}\boldsymbol{m}^T\}$ to form $\hat{\theta}$. Let $\mathcal{D} = \mathbb{E}\{(\theta - \hat{\theta})^2\}$ denote the MSE corresponding to the LMMSE estimator, where $\mathcal{D}$ depends on $r_k \ \forall k$. We consider two system-level constrained optimization problems with respect to optimization variables $r_k \ \forall k$. In the first problem, we minimize $\mathcal{D}$ subject to the network bandwidth constraint. In the second problem, we minimize the total number of transmitted bits subject to the constraint on $\mathcal{D}$. In other words, we are interested to solve the following

---

[1] For the general case $\mathbb{E}\{h_k\} = \mu_k$ we can scale $x_k$ and obtain $x'_k = h'_k \theta + n'_k$, where $x'_k = x_k/\mu_k$, $h'_k = h_k/\mu_k$, $n'_k = n_k/\mu_k$, $\mathbb{E}\{h'_k\} = 1$, $\text{var}(h'_k) = \sigma_{h_k}^2/\mu_k^2$, $n'_k \sim \mathcal{N}(0, \sigma_{n_k}^2/\mu_k^2)$. Thus without loss of generality, we assume $\mathbb{E}\{h_k\} = 1, \forall k$.

two constrained optimization problems:

$$(\textbf{P1}) \quad \underset{r_k \ \forall k}{\text{minimize}} \quad \mathcal{D}(\{r_k\}_{k=1}^K) \quad (2)$$
$$\text{s.t.} \sum_{k=1}^K r_k \leq B_{tot}, r_k \in \mathbb{Z}_+, \ \forall k,$$

$$(\textbf{P2}) \quad \underset{r_k \ \forall k}{\text{minimize}} \quad \sum_{k=1}^K r_k \quad (3)$$
$$\text{s.t.} \quad \mathcal{D}(\{r_k\}_{k=1}^K) \leq \mathcal{D}_0, \ r_k \in \mathbb{Z}_+, \ \forall k,$$

where $\mathcal{D}_0$ is the pre-determined upper bound on $\mathcal{D}$.

## III. CHARACTERIZING MSE FOR LMMSE ESTIMATOR

We wish to characterize $\mathcal{D}$ in terms of the optimization variables $\{r_k\}_{k=1}^K$. From [36] we have:

$$\hat{\theta} = \boldsymbol{Gm} \quad \text{where} \quad \boldsymbol{G} = \mathbb{E}\{\theta \boldsymbol{m}^T\}(\mathbb{E}\{\boldsymbol{mm}^T\})^{-1}, \quad (4)$$

$$\mathcal{D} = \sigma_\theta^2 - \mathbb{E}\{\theta \boldsymbol{m}^T\}(\mathbb{E}\{\boldsymbol{mm}^T\})^{-1}\mathbb{E}\{\theta \boldsymbol{m}^T\}^T). \quad (5)$$

To find $\mathbb{E}\{\theta \boldsymbol{m}^T\}$ and $\mathbb{E}\{\boldsymbol{mm}^T\}$ in (4), (5) we need to delve into statistics of the quantization errors.

• **Characterizing $\mathbb{E}\{\theta \boldsymbol{m}^T\}$ and $\mathbb{E}\{\boldsymbol{mm}^T\}$**: For sensor $k$, let the difference between observation $x_k$ and its quantized version $m_k$, i.e., $\epsilon_k = x_k - m_k$, be the corresponding quantization error. In general, $\epsilon_k$'s are mutually correlated and also are correlated with $x_k$'s. However, in [37] it is shown that, when highly correlated Gaussian random variables are coarsely quantized with uniform quantizers of step sizes $\Delta_k$'s, quantization errors can be approximated as mutually independent random variables, that are uniformly distributed in the interval $[-\frac{\Delta_k}{2}, \frac{\Delta_k}{2}]$, and are also independent of quantizer inputs. Here, since $\theta$ and $n_k$'s in (1) are Gaussian, conditioned on $h_k$'s observations $x_k$'s are correlated Gaussian that are quantized with uniform quantizers of quantization step sizes $\Delta_k$'s. Thus $\epsilon_k$'s are approximated as mutually independent zero mean uniform random variables with variance $\sigma_{\epsilon_k}^2 = \frac{\Delta_k^2}{12}$, that are also independent of $x_k$'s (and hence independent of $\theta$ and $n_k$'s). Using the aforementioned assumptions and approximations for the quantization errors, $k$th element of $\mathbb{E}\{\theta \boldsymbol{m}^T\}$ becomes:

$$\mathbb{E}\{\theta m_k\} = \mathbb{E}_{h_k}\{\mathbb{E}\{\theta m_k | h_k\}\} = \mathbb{E}_{h_k}\{\mathbb{E}\{\theta(x_k - \epsilon_k)|h_k\}\}$$
$$= \mathbb{E}_{h_k}\{h_k \mathbb{E}\{\theta^2\} + \mathbb{E}\{n_k\}\mathbb{E}\{\theta\} - \mathbb{E}\{\epsilon_k\}\mathbb{E}\{\theta\}\} = \sigma_\theta^2. \quad (6)$$

Hence we have $\mathbb{E}\{\theta \boldsymbol{m}^T\} = \sigma_\theta^2 \boldsymbol{1}^T$, where $\boldsymbol{1} = [1, ..., 1]^T$. Similarly, for $(k,l)$th element of $\mathbb{E}\{\boldsymbol{mm}^T\}$ we have:

$$\mathbb{E}\{m_k m_l\} = \mathbb{E}_{h_k, h_l}\{\mathbb{E}\{m_k m_l | h_k, h_l\}\} = \quad (7)$$
$$\mathbb{E}_{h_k, h_l}\{\mathbb{E}\{(h_k \theta + n_k - \epsilon_k)(h_l \theta + n_l - \epsilon_l)|h_k, h_l\}\} \stackrel{(a)}{=}$$
$$\mathbb{E}_{h_k, h_l}\{h_k h_l \mathbb{E}\{\theta^2\} + \mathbb{E}\{n_k n_l\} + \mathbb{E}\{\epsilon_k \epsilon_l\}\} =$$
$$\mathbb{E}_{h_k, h_l}\{h_k h_l\}\mathbb{E}\{\theta^2\} + \mathbb{E}\{n_k n_l\} + \mathbb{E}\{\epsilon_k \epsilon_l\},$$

where for (a) we have used the assumptions that (i) $n_k$'s and $\theta$ are uncorrelated, (ii) $\epsilon_k$'s and $\theta$ are uncorrelated (iii) $n_k$'s and $\epsilon_k$'s are uncorrelated. Having (7), and noting that $h_k$'s are uncorrelated with unit means, we reach:

$$\mathbb{E}\{m_k m_l\} = \begin{cases} \sigma_\theta^2 + \sigma_k^2 + \sigma_{\epsilon_k}^2, & \text{if} \quad k = l \\ \sigma_\theta^2, & \text{if} \quad k \neq l \end{cases}$$

where $\sigma_k^2 = \sigma_\theta^2 \sigma_{h_k}^2 + \sigma_{n_k}^2$. Consequently matrix $\mathbb{E}\{\boldsymbol{mm}^T\}$ can be written as the following:

$$\mathbb{E}\{\boldsymbol{mm^T}\} = \sigma_\theta^2 \boldsymbol{1}\boldsymbol{1}^T + \text{diag}(\frac{1}{\alpha_1}, ..., \frac{1}{\alpha_K}), \quad (8)$$

where $\alpha_k^{-1} = \sigma_k^2 + \sigma_{\epsilon_k}^2$. Applying matrix inversion Lemma [36] to (8) we find:

$$[(\mathbb{E}\{\boldsymbol{mm}^T\})^{-1}]_{k,l} = \begin{cases} \alpha_k - \frac{\alpha_k^2}{\sigma_\theta^{-2} + \sum_{k=1}^K \alpha_k}, & \text{if} \quad k = l \\ -\frac{\alpha_k \alpha_l}{\sigma_\theta^{-2} + \sum_{k=1}^K \alpha_k}, & \text{if} \quad k \neq l \end{cases}$$

Proposition 1 summarizes the expressions for $\hat{\theta}, \mathcal{D}$ in (4),(5).

**Proposition 1.** *The LMMSE estimator $\hat{\theta}$ and its corresponding MSE $\mathcal{D}$, based on the quantized observations $\{m_k\}_{k=1}^K$ are:*

$$\hat{\theta} = \sum_{k=1}^K c_k m_k \quad \text{where} \quad c_k = \frac{\alpha_k}{\sigma_\theta^{-2} + \sum_{k=1}^K \alpha_k},$$

$$\mathcal{D} = \frac{1}{\sigma_\theta^{-2} + \sum_{k=1}^K \alpha_k}. \quad (9)$$

Examining (9), we note that $\alpha_k$ represents the contribution of sensor $k$ in reducing the overall MSE at the FC. Also, $\alpha_k$ can be viewed as an indicator for the quality of received message from sensor $k$: the larger $\alpha_k$ is, the more reliable is the received message. It is easy to verify that $\alpha_k$ is increasing in $r_k$ and decreasing in $\sigma_k^2$.

**Remark 1:** When all observations $x_k$'s are available at the FC with full precision (so-called centralized estimation), the LMMSE estimator would be $\breve{\theta} = \sum_{k=1}^K b_k x_k$, where $b_k = \frac{\sigma_k^{-2}}{\sigma_\theta^{-2} + \sum_{k=1}^K \sigma_k^{-2}}$, with its corresponding MSE $\mathcal{D}^c = \frac{1}{\sigma_\theta^{-2} + \sum_{k=1}^K \sigma_k^{-2}}$. This clairvoyant estimator can be used as our performance benchmark, since $\mathcal{D} > \mathcal{D}^c$.

**Proposition 2.** *In a network with homogeneous sensors, i.e., $\sigma_k^2 = \sigma^2, \forall k$, and all sensors quantize their observations with identical quantizers of step size $\Delta$, the MSE gap between two linear estimators $\hat{\theta}$ and $\breve{\theta}$ is:*

$$\mathcal{D} - \mathcal{D}^c = \frac{K\Delta^2}{12(K + \sigma_\theta^{-2}(\sigma^2 + \sigma_\epsilon^2))(K + \sigma_\theta^{-2}\sigma^2)} \leq \frac{\Delta^2}{12K}. \quad (10)$$

Based on (10), if $\Delta \to 0$, then $\mathcal{D} \to \mathcal{D}^c$. Additionally, if $K \to \infty$, then $\mathcal{D} \to \mathcal{D}^c$ even for large $\Delta$. These conclusions still hold true in a network with heterogeneous sensors, where sensor $k$ quantizes with step size of $\Delta_k$. If $(\max \Delta_k) \to 0$, then $\alpha_k \to \sigma_k^{-2}$ and according to (9), $\mathcal{D} \to \mathcal{D}^c$. On the other hand, according to (9) and noting that $\alpha_k > 0$ for active sensors, MSE always decreases as the number of active sensors increases. Thus as $K \to \infty$, we have $\mathcal{D} \to \mathcal{D}^c \to 0$.

## IV. SOLVING CONSTRAINED PROBLEM (P1)

Since the optimization variables $r_k$'s are integer and $\mathcal{D}$ is a non-linear function of $r_k$'s, (**P1**) is a non-linear integer programing problem and is NP-hard. Even if the inequality constrain holds with equality, i.e., $\sum_{k=1}^K r_k = B_{tot}$, solving (**P1**) requires a brute-force evaluation over $\binom{K+B_{tot}-1}{K-1}$ choices. For $K = 50$ and $B_{tot} = 60$ bits, the number of evaluations would be in the order of $10^{31}$. The following lemmas help us find strategies that reduce the computational complexity required for solving (**P1**).

**Lemma 1.** *Minimizing $\mathcal{D}$ in (9), is equivalent to maximizing $\sum_{k=1}^{K} \alpha_k$ with the same constraints as in (2).*

*Proof.* Since $\sigma_\theta^{-2} > 0$, it is axiomatic. □

**Lemma 2.** *Suppose $\{r_k^*\}_{k=1}^{K}$ is the optimal solution to (**P1**). Then $\sum_{k=1}^{K} r_k^* = B_{tot}$.*

*Proof.* Note $\alpha_k$ is a function of $r_k$ through $\sigma_{\epsilon_k}^2 = \frac{\Delta_k^2}{12} = \frac{\tau_k^2}{3(2^{r_k}-1)^2}$. It is easy to verify that $\frac{\partial \mathcal{D}}{\partial r_k} \leq 0$ and $\mathcal{D}$ is a decreasing function of $r_k$'s. Thus the optimal solution satisfies the network bandwidth constraint, i.e., $\sum_{k=1}^{K} r_k^* = B_{tot}$. □

**Lemma 3.** *Without loss of generality, suppose sensors are sorted[2] such that $\sigma_1^2 \leq \sigma_2^2 \leq ... \leq \sigma_K^2$. Then the optimal solution satisfies $r_i^* \geq r_j^*$ for $i < j$.*

*Proof.* Suppose $\{r_k\}_{k=1}^{K}$ is the optimal solution, such that $r_i < r_j$ for $i < j$. Also, suppose $\{r_k'\}_{k=1}^{K}$ is a solution of (**P1**), which is not necessarily optimal, such that $r_k' = r_k$ for $k \neq i,j$ and $r_i' = r_j, r_j' = r_i$. Consider the following:

$$\sum_{k=1}^{K} \alpha_k(r_k') - \sum_{k=1}^{K} \alpha_k(r_k) = \overbrace{\sum_{k\neq i,j}^{K} \alpha_k(r_k) - \sum_{\neq i,j}^{K} \alpha_k(r_k)}^{=\delta_1}$$

$$+ \overbrace{\frac{1}{\sigma_i^2 + \frac{\tau^2}{3(2^{r_j}-1)^2}} - \frac{1}{\sigma_i^2 + \frac{\tau^2}{3(2^{r_i}-1)^2}}}^{=\delta_2}$$

$$+ \overbrace{\frac{1}{\sigma_j^2 + \frac{\tau^2}{3(2^{r_j}-1)^2}} - \frac{1}{\sigma_j^2 + \frac{\tau^2}{3(2^{r_i}-1)^2}}}^{=\delta_3} > 0 \ .$$

One can verify $\delta_1 = 0, \delta_2 > 0, \delta_3 > 0$, thus $\sum_{k=1}^{K} \alpha_k(r_k') > \sum_{k=1}^{K} \alpha_k(r_k)$. According to (9), the MSE associated with $\{r_k'\}_{k=1}^{K}$ should be less than that of the optimal solution $\{r_k\}_{k=1}^{K}$, which is a contradiction. In this proof, we assumed $\tau_k = \tau, \forall k$, although the proof is still valid for unequal $\tau_k$'s, provided that $\tau_i \leq \tau_j$, which is satisfied if we choose $\tau_k^2 \propto \text{var}(x_k) = \sigma_\theta^2 + \sigma_k^2$. □

In the next subsections, we propose four methods for solving (**P1**): *A)* longest root to leaf path method, which is optimal with less computational complexity than that of brute force, *B)* greedy method, *C)* integer relaxation method, *D)* individual rate allocation method. The suboptimal *B)*, *C)*, *D)* methods have moderate to low computational complexity.

### A. Longest Root to Leaf Path Method

We view (**P1**) as the problem of finding the longest "root to leaf" path in a weighted directed binary tree, where there is a constraint on the number of edges from "root to leaf" [38]. In fact our objective function $\sum_{k=1}^{K} \alpha_k$ can be viewed as the length of the path to be maximized, where the constraint on the number of edges is $\sum_{k=1}^{K} r_k \leq B_{tot}$. Fig. 1 demonstrates the problem for $K = 3$ and $B_{tot} = 5$ bits. The nodes are tagged with indices of sorted sensors and visiting node $k$ is translated to "allocating one bit to sensor $k$". The edge[3] weight $w_k(r)$ is the weight of the edge entering node $k$ and $r$ is the number of prior visits of node $k$, i.e., $w_k(r) = \alpha_k(r+1) - \alpha_k(r) = \alpha_k$(number of bits allocated to sensor $k$ so far $+1$) $- \alpha_k$(number of bits allocated to sensor $k$ so far). For instance, the green path in Fig. 1 is associated with the rate allocation $[r_1, r_2, r_3] = [3, 2, 0]$, and the corresponding objective function value is $\sum_{k=1}^{K} \alpha_k = w_1(1) + w_1(2) + w_2(0) + w_2(1) = \alpha_1(3) + \alpha_2(2)$. To solve (**P1**), one needs to construct the associated binary tree with structures conforming to lemmas 2 and 3, then uses a search algorithm, such as depth first search (DFS) [38] to discover all possible "root to leaf" paths, and choose the path that results in the maximum objective function value. For $K = 3$ and $B_{tot} = 5$ bits, Fig. 1 shows there exist 5 "root to leaf" paths, all conforming to lemmas 2 and 3, corresponding to 5 distinct rate allocation among 3 sensors $[r_1, r_2, r_3] \in \{[5,0,0], [4,1,0], [3,2,0], [3,1,1], [2,2,1]\}$. We recognize these as different partitions of the integer number 5, with 3 or fewer addends [39], i.e., the number of possible "root to leaf" paths in a binary tree constructed as explained, conforming to lemmas 2 and 3, and characterizing (**P1**), is equal to the number of solutions to the following equation:

$$r_1 + r_2 + ... + r_K = B_{tot} \qquad (11)$$
$$\text{s.t.} \quad r_1 \geq r_2 \geq ... \geq r_K \quad r_k \in \mathbb{Z}_+ \ .$$

---
[2] We assume sorted sensors throughout this work for all scenarios.

Although the number of ways one can partition an integer number does not have a closed form formula, the literature [39] provides some useful asymptotic formulas or recurrence relations. Suppose $q_k(n)$ is the number of solutions to (11), then we have the recurrence relation $q_k(n) = q_{k-1}(n) - q_k(n-k)$, with $q_0(n) = 0$, $q_k(1) = 1$ [39]. For $K = 50$ and $B_{tot} = 60$ bits, $q_K(B_{tot})$ is in the order of $10^6$, which is much smaller than that of brute force $10^{31}$. The computational complexity of this method is still high for very large networks, e.g., $K \geq 100$, and hence its application is most beneficial for finding the optimal solution of small to moderate size networks.

### B. Greedy Method

Recall from Lemma 1 that the maximum reduction in $\mathcal{D}$ corresponds to the maximum increase in $\sum_{k=1}^{K} \alpha_k(r_k)$. Hence, our proposed greedy method in each iteration allocates one bit to the sensor that guarantees the maximum increase in $\sum_{k=1}^{K} \alpha_k(r_k)$, i.e., in each iteration the algorithm loads one bit on sensor $k^*$ where $k^* = \underset{k}{\text{argmax}}\ I_k(r_k) = \underset{k}{\text{argmax}}(\alpha_k(r_k + 1) - \alpha_k(r_k))$. The iteration ends when all $B_{tot}$ bits are allocated to the sensors. Following algorithm illustrates the details:

For $K = 3$ and $B_{tot} = 5$ bits Fig. 1 shows the accepted decisions by the greedy method at each iteration/decision epoch with green arrows and the rejected decisions with red arrows. Note that the initial point is always $r_k = 1, r_k = 0$ for $k = 2, ..., K$, since the first bit is always allocated to sensor 1 (for sorted sensors sensor 1 has the largest $\alpha_k$ or smallest $\sigma_k^2$). The second bit can be allocated to either sensor 1 or sensor 2,

---
[3] For definition of weight $w_k(0)$, we consider $\alpha_k(0) = 0$.

**Data:** $B_{tot}, \{\tau_k\}_{k=1}^K, \{\sigma_k^2\}_{k=1}^K$
**Result:** rate allocation $\{r_k^*\}_{k=1}^K$
initialization;
$r_1 = 1, r_k = 0$ for $k = 2, ..., K$, $\mathcal{S} = \{1, 2\}$
**for** $i = 1 : B_{tot}$ **do**
$\quad k^* = \underset{k \in \mathcal{S}}{\text{argmax}}(\alpha_k(r_k + 1) - \alpha_k(r_k))$
$\quad r_{k^*} = r_{k^*} + 1$
$\quad \mathcal{S} = \{k | r_k < r_{k-1}\} \cup \{1\}$
**end**
**Algorithm:** greedy method for rate allocation in (**P1**)

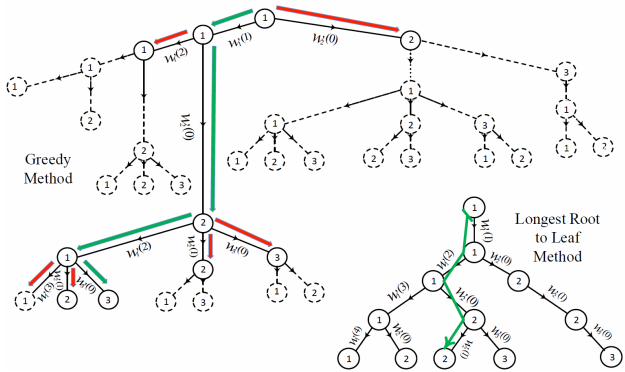

Fig. 1: Longest root to leaf path and Greedy method

i.e., $k^* = \underset{k \in \{1,2\}}{\text{argmax}}(\alpha_k(r_k + 1) - \alpha_k(r_k))$, this is equivalent to making the decision $w_1(1) \underset{k^*=2}{\overset{k^*=1}{\gtrless}} w_2(0)$ (look at the weights on the edges in Fig. 1). The sequence of green arrows in Fig.1, $1 \to 1 \to 2 \to 1 \to 3$ is associated with the rate allocation $[r_1, r_2, r_3] = [3, 1, 1]$.

In the following, we look at the computational complexity of the greedy method in two cases: case *(a)* $B_{tot} \leq K$, in this case the first bit has to be allocated to sensor 1, the second bit can be allocated to either sensor 1 or sensor 2. In general the $i$th bit, for $1 \leq i \leq B_{tot}$, can be allocated to one of at most $i$ sensors (sensor 1,..., sensor $i$). In other words, in the $i$th decision epoch, greedy method should find the best sensor among eligible candidates in set $\mathcal{S} = \{k | k \leq i, r_k < r_{k-1}\} \cup \{1\}$, where $\mathcal{S}$ has a maximum of $i$ elements. Thus allocating $B_{tot}$ bits among $K$ sensors needs calculation of $I_k(r_k) = \alpha_k(r_k + 1) - \alpha_k(r_k)$ for at most $2 + 3 + ... + B_{tot} = (\frac{B_{tot}^2 + B_{tot}}{2}) - 1$ times. Case *(b)* $B_{tot} > K$, adopting a reasoning similar to case *(a)*, in this case allocation of the first $K$ bits needs at most $(\frac{K^2+K}{2}) - 1$ calculations. Each of the remaining $B_{tot} - K$ bits can be allocated to one of at most $K$ sensors, leading into $(B_{tot} - K) \times K$ number of calculations at most. Hence, the overall number of evaluations at most would be $(\frac{K^2+K}{2}) - 1 + (B_{tot} - K) \times K \approx K(B_{tot} - K/2)$. For $K = 50$ and $B_{tot} = 60$ bits, where $B_{tot} > K$, the number of evaluations would be in order of $10^3$.

**Remark 2:** In the absence of a powerful FC, the proposed greedy algorithm can be implemented in a distributed way, assuming sensors can broadcast and hear the broadcast messages by other sensors. Sensor $k$ calculates the value $I_k(r_k) = \alpha_k(r_k + 1) - \alpha_k(r_k)$, and broadcasts the value. Hearing all $I_j, j \neq k$, sensor $k$ increases $r_k$ by one if it has the largest $I_k$ among all sensors. Doing the mentioned process for $B_{tot}$ times would complete the rate allocation.

*C. Integer Relaxation Method*

Convex relaxation for solving combinatorial optimization problems such as (**P1**) is a rather old technique, that has been widely used in research and applied to a variety of applications [40], [41]. Relaxing the integer constraint on $r_k$'s and letting them be positive numbers and using lemmas 1 and 2, we consider the following relaxed problem:

$$(\mathbf{P'1}) \quad \underset{r_k \, \forall k}{\text{maximize}} \sum_{k=1}^K \alpha_k(r_k) \quad (12)$$

$$\text{s.t.} \sum_{k=1}^K r_k = B_{tot}, r_k \in \mathbb{R}_+, \, \forall k.$$

The Lagrangian for (**P'1**) is:

$$\mathcal{L}(\{r_k, \mu_k\}_{k=1}^K, \lambda) = \sum_{k=1}^K \alpha_k(r_k) - \mu_k r_k + \lambda(\sum_{k=1}^K r_k - B_{tot}).$$

In the following we apply the first order KKT necessary optimality conditions for (**P'1**) which generate a closed-form solution for $r_k$'s. Afterwards, we prove that the obtained solution satisfies the second order sufficient optimality conditions.

• *Necessary Optimality Conditions*

After solving the KKT conditions corresponding to (12), we find:

$$r_k^\dagger = 0.5 \left[ \log_2 \left( \tau_k^2 (\lambda^\dagger - \sigma_k^2 - \sqrt{\lambda^{\dagger 2} - 2\lambda^\dagger \sigma_k^2})^{-1} \right) - \log_2 3 \right]^+ \quad (13)$$

where $[x]^+ = \max(0, x)$ and $\lambda^\dagger$ in (13) is the solution to following equation:

$$g(\lambda, \{\sigma_k^2, \tau_k^2\}_{k=1}^{K^\dagger}) = \prod_{k=1}^{K^\dagger} \tau_k^{-2}(\lambda - \sigma_k^2 - \sqrt{\lambda^2 - 2\lambda\sigma_k^2}) = T, \quad (14)$$

in which $K^\dagger = \max\{k | \lambda^\dagger > 2\sigma_{K^\dagger}^2, r_k^\dagger > 0\}$, and $T = 4^{-B_{tot}} 3^{-K^\dagger}$. Consider a new equation which is obtained by replacing $K^\dagger$ in (14) with $M$. The new equation, which we refer to as (14'), does not necessarily have a *real* solution for $\lambda$, such that $\lambda > 2\sigma_M^2$ for any value of $M \in \{2, 3, ..., K\}$. In order to find the requirements for (14') to yield a real solution for $\lambda$, we present the following Lemma and ensuing discussion. For simplicity, we drop the parameters $\{\sigma_k^2, \tau_k^2\}_{k=1}^M$ in $g(\lambda, \{\sigma_k^2, \tau_k^2\}_{k=1}^M)$ and indicate it as $g(\lambda, M)$.

**Lemma 4.** *The function $g(\lambda, M)$ is a decreasing function of $\lambda$, for $\lambda > 2\sigma_M^2$.*

*Proof.* Consider $g(\lambda, M) = \prod_{k=1}^M g_k(\lambda)$, where $g_k(\lambda) = \tau_k^{-2}(\lambda - \sigma_k^2 - \sqrt{\lambda^2 - 2\lambda\sigma_k^2})$. We can verify that $g_k(\lambda)$'s are *strictly* decreasing in $\lambda$, because $\frac{dg_k(\lambda)}{d\lambda} = 1 - \frac{\lambda - \sigma_k^2}{\sqrt{\lambda^2 - 2\lambda\sigma_k^2}} < 0$. Since all $g_k(\lambda)$'s are *strictly* decreasing and positive, i.e., $g_k(\lambda) > 0$, $\forall k$, we conclude that $g(\lambda, M)$ is a *strictly* decreasing function in $\lambda$. □

Having Lemma 4, we consider two scenarios that occur when solving $g(\lambda, M) = T$: case *(i)* when $T \leq g(\lambda)|_{\lambda=\sigma_M^2}$, in this case according to Lemma 4 we have a unique *real* solution for $\lambda$; case *(ii)* when $T > g(\lambda)|_{\lambda=\sigma_M^2}$, in this case there is no *real* solution for $\lambda$. Hence we need to increase the value of $g(\lambda)|_{\lambda=\sigma_M^2}$ to reach $T$. The only way to accomplish this is decreasing the number of active sensors that contribute to $g(\lambda)|_{\lambda=\sigma_M^2}$ and deactivating sensors with largest $\sigma_k^2$ values, until we find a *real* solution for $\lambda$ or only one active sensor remains. In other words, solving $(14')$ in case *(ii)* translates into obtaining the set of active sensors $\mathcal{A} = \{1, 2, ..., K^\dagger\}$ and allocating $B_{tot}$ among these active sensors.

**Remark 3**: The solution in (13) can be implemented in a distributed fashion. FC solves (14) and broadcasts $\lambda^\dagger$. Each sensor calculates its own $r_k^\dagger$ using $\lambda^\dagger$ via (13). If a sensor finds its rate to be zero or a *non-real* value, it means that the sensor must be inactive. The integer relaxation method has a very low computational complexity, since it requires finding the root of the monotonic function in (14) once and, and then calculating the rates via (13) for a maximum of $K$ times.

We can find an approximate closed form solution for (14) under the special condition when $(\lambda - \sigma_k^2)^2$ is large compared to $\sigma_k^4$. Rewriting the function $g_k(\lambda) = \tau_k^{-2}[\lambda - \sigma_k^2 - ((\lambda - \sigma_k^2)^2 - \sigma_k^4)^{1/2}]$ and keeping only the first two terms in the binomial expansion of the term $((\lambda-\sigma_k^2)^2-\sigma_k^4)^{1/2}$, we obtain $g_k(\lambda) \approx \frac{\sigma_k^4}{2\tau_k^2(\lambda-\sigma_k^2)} \approx \frac{\sigma_k^4}{2\tau_k^2\lambda}$. Substituting the approximation in (14), we reach $\prod_k g_k(\lambda) \approx \prod_k 2^{-1}\sigma_k^4 \tau_k^{-2} \lambda^{-1}$, based on which the Lagrange multiplier can be approximated as $\lambda^\dagger \approx 1.5 \ \eta^2 \ 4^{\frac{B_{tot}}{K^\dagger}}$, where $\eta = \prod_k \sigma_k^2 \tau_k^{-1}$. Substituting the approximation for $\lambda^\dagger$ in (13) gives the following:

$$r_k^\dagger \approx \left[\frac{B_{tot}}{K^\dagger} + \log_2(\eta \frac{\tau_k}{\sigma_k^2})\right]^+. \quad (15)$$

Examining (15), we note that first term inside the bracket is common among active sensors and can be perceived as *average rate*, whereas the second term (which depends on $\tau_k$, $\sigma_k^2$) differs among active sensors, such that an active sensor with a larger ratio $\frac{\tau_k}{\sigma_k^2}$ is allocated a larger $r_k^\dagger$. Consistent with the assumption in the proof of Lemma 3, suppose $\tau_k = \kappa \text{var}(x_k) = \kappa(\sigma_\theta^2 + \sigma_k^2)$. Interestingly, the second term in (15) takes the form $\log_2(\kappa\eta(1 + \frac{\sigma_\theta^2}{\sigma_k^2}))$, where the ratio $\frac{\sigma_\theta^2}{\sigma_k^2}$, can be viewed as the observation *SNR* in (1). We consider two scenarios: *(i) high observation SNR*: the quantization rates are large (fine quantization) and less sensors become active for a given $B_{tot}$. *(ii) low observation SNR*: the quantization rates are smaller (coarse quantization) and more sensors become active for the same $B_{tot}$ value, compared with that of scenario *(i)*. Substituting (15) in (9) and after some simplifications we establish the bound $\mathcal{D} \leq \mathcal{D}^c(1 + \frac{\sigma_{K^\dagger}^2 4^{-\frac{B_{tot}}{K^\dagger}}}{3\eta^2})$.

• *Sufficient Optimality Conditions*

The objective and equality constraint functions in (12) are twice differentiable. Hence, the second order sufficient optimality conditions for the solution in (13) and (14) to be a strict minimum for $(\mathbf{P'1})$ are [42, p.301, proposition 3.2.1]:

$$-\boldsymbol{y}^T \left(\nabla_{\boldsymbol{r}}^2 \mathcal{L}(\{r_k, \mu_k\}_{k=1}^K, \lambda)|_{\substack{\lambda=\lambda^\dagger \\ r_k=r_k^\dagger \ \forall k}}\right) \boldsymbol{y} > 0, \quad \forall \boldsymbol{y} \neq 0$$
$$\text{with} \ [\nabla(\sum_{k \in \mathcal{A}} r_k^\dagger - B_{tot})]^T \boldsymbol{y} = 0, \quad (16)$$

where $\nabla_{\boldsymbol{r}}^2 \mathcal{L}(\{r_k, \mu_k\}_{k=1}^K, \lambda)$ is the Hessian matrix of the Lagrangian in (13), and $\nabla(\sum_{k\in\mathcal{A}} r_k^\dagger - B_{tot})$ is the gradient of the equality constraint in (12), both evaluated at the solution in (13) and (14). It is easy to verify that the Hessian matrix is diagonal with entries:

$$[\nabla_{\boldsymbol{r}}^2\mathcal{L}(\{r_k,\mu_k\}_{k=1}^K,\lambda)]_{k,k} = \sigma_{\epsilon_k}^2(\ln 4)^2 \overbrace{\frac{\sigma_{\epsilon_k}^2 - \sigma_k^2}{(\sigma_k^2+\sigma_{\epsilon_k}^2)^3}}^{=\beta}, \forall k. \quad (17)$$

Noting that the denominator in (17) and $\sigma_{\epsilon_k}^2(\ln 4)^2$ are positive numbers we probe into $\beta$ evaluated at the solution in (13):

$$\beta^\dagger = (\sigma_{\epsilon_k}^2 - \sigma_k^2) = \lambda^\dagger - 2\sigma_k^2 - \sqrt{\lambda^{\dagger 2} - 2\lambda^\dagger \sigma_k^2} < 0. \quad (18)$$

The inequality in (18) is true, because $\beta$ (which is a function of $\lambda^\dagger$) is decreasing in $\lambda^\dagger$ and noting that $\lambda^\dagger > 2\sigma_{K^\dagger}^2$ in (18), we have $\sup_{k,\lambda^\dagger}(\lambda^\dagger - 2\sigma_k^2 - \sqrt{\lambda^{\dagger 2} - 2\lambda^\dagger \sigma_k^2}) = 0$. Therefore $\left[\nabla_{\boldsymbol{r}}^2\mathcal{L}(\{r_k,\mu_k\}_{k=1}^K,\lambda)|_{\substack{\lambda=\lambda^\dagger \\ r_k=r_k^\dagger \ \forall k}}\right]_{k,k} < 0 \ \forall k$, confirming that the sufficient optimality conditions in (16) are satisfied.

• *Migration to Integer Solution*

We describe an approach for migrating from the continuous solution in (13) to an integer solution satisfying the integer constraint in (2). We round the rates to nearest integers [43], [44]. In case the rounding violates the bandwidth constraint, we reduce the smallest rate by one, because this sensor is more likely to be the weakest player in the network (in the sense that it has the least contribution to $\mathcal{D}$) until the bandwidth constraint is satisfied. Although rounding the rates to nearest integers may sounds trivial, our simulation results corroborate that the performance loss is negligible, while at the same time it keeps the rate allocation scheme simple and easily implementable.

### D. Individual Rate Allocation Method

Examining (10) closely we realize that allocating $B_{tot}$ among sensors in order to minimize $\mathcal{D}$ presents a trade-off between the number of active sensors and quantization accuracy. If $B_{tot}$ is distributed among only few sensors, we can have fine quantization, i.e., small $K$ and small $\Delta$. On the other hand, if $B_{tot}$ is distributed among many sensors, we can only have coarse quantization, i.e., large $K$ and large $\Delta$. Consider a network with homogeneous sensors $\sigma_k^2 = \sigma^2, \forall k$. Given $B_{tot}$, there exists an optimal number of active sensors $K^{opt}$, associated with an optimal quantization rate $r^{opt}$, where $K^{opt} r^{opt} = B_{tot}$. Thus the maximization of $\sum_{k=1}^K \alpha_k = K\alpha$, where we substitute $K = B_{tot}/r$, reduces to the following one dimensional simple search for $r^{opt}$ in the set $\mathcal{S}_h = \{1, ..., B_{tot}\}$:

$$r^{opt} = \underset{r \in \mathcal{S}_h}{\arg\min}\{r(\sigma^2 + \tau^2 3^{-1}(2^r - 1)^{-2})\}, \quad (19)$$

and consequently $K^{opt} = \lfloor \frac{B_{tot}}{r^{opt}} \rfloor$. Modifying the solution in (19) for heterogeneous networks, we reach the following:

$$r_k^{sopt} = \underset{r \in \mathcal{S}_k}{\operatorname{argmin}}\{r(\sigma_k^2 + \tau_k^2 3^{-1}(2^r-1)^{-2})\}, \quad (20)$$

in which $\mathcal{S}_1 = \{1, ..., B_{tot}\}$, $\mathcal{S}_k = \{1, ..., B_{tot} - \sum_{i=1}^{k-1} r_i^{sopt}\}$ for $k = 2, ..., K$, $K^{sopt} = \max\{k| \mathcal{S}_k \neq \{0\}\}$, $r_k^{sopt}$ and $K^{sopt}$ are the rates and number of active sensors, respectively. Note that the solution in (20) is integer and unique (since the objective function in (20) is convex for $r > 0$). The drawback of the proposed rate allocation method is that for large $B_{tot}$, all $B_{tot}$ bits may not be allocated to sensors, i.e., $\sum_{k=1}^{K^{sopt}} r_k^{sopt} < B_{tot}$, causing the solution in (20) to deviate from the optimal solution according to Lemma 2. This method is similar to the one in [6], with the difference that starting from sensor 1, we update and reduce the search domain, i.e., $\mathcal{S}_k$ for the next sensor. This accelerates the rate allocation process. Additionally, search domain reduction in some scenarios would help to use all $B_{tot}$ bits by activating more sensors with coarse quantizers though. The proposed method exhibits a moderate computational complexity, since it only requires solving (20) for a maximum of $K$ times and it is *almost fully distributed* [6].

## V. SOLVING CONSTRAINED PROBLEM (P2)

Different from (**P1**), satisfying the MSE constraint $\mathcal{D}(\{r_k\}_{k=1}^K) \leq \mathcal{D}_0$ in (**P2**) enforces the number of active sensors to exceed a minimum number $K_{min}$. Lemma 5 provides $K_{min}$.

**Lemma 5.** *To satisfy the the constraint $\mathcal{D}(\{r_k\}_{k=1}^K) \leq \mathcal{D}_0$ we need at least $K_{min}$ active sensors, where $K_{min} = \min\{K| \sum_{k=1}^K \sigma_k^{-2} > \mathcal{D}_0' = \mathcal{D}_0^{-1} - \sigma_\theta^{-2}\}$.*

*Proof.* Considering (9) and the definition of $\alpha_k$'s, we find that $\mathcal{D}(\{r_k\}_{k=1}^K) \leq \mathcal{D}_0$ is equivalent to $\sum_{k=1}^K \alpha_k \geq \mathcal{D}_0' = \mathcal{D}_0^{-1} - \sigma_\theta^{-2}$. Thus (**P2**) is equivalent to minimizing $\sum_{k=1}^K r_k$ such that $\sum_{k=1}^K \alpha_k \geq \mathcal{D}_0'$ and $r_k \in \mathbb{Z}_+$, $\forall k$. Since $\alpha_k > 0$, $\forall k$, we can increase $\sum_{k=1}^K \alpha_k = \sum_{k=1}^K (\frac{1}{\sigma_k^2 + \sigma_{\epsilon_k}^2})$, via increasing the number of active sensors, until the MSE constraint is satisfied. This implies that the minimum number of active sensors can be found by letting $\sigma_{\epsilon_k}^2 = 0, \forall k$, i.e., $K_{min} = \min\{K| \sum_{k=1}^K \sigma_k^{-2} > \mathcal{D}_0'\}$. □

In the following we propose three methods for solving (**P2**): *A*) greedy method, *B*) integer relaxation method, *C*) individual rate allocation method. We obtain these methods via applying some modifications to the proposed methods in section IV.

### A. Greedy Method

According to Lemma 5, we need at least $K_{min}$ active sensors. Therefore, we initiate the algorithm with $r_k = 1$ for $k \in \{1, ..., K_{min}\}$ and let $r_k = 0$ otherwise, and go through the greedy method until the MSE constraint is satisfied.

**Data:** $\mathcal{D}_0, \sigma_\theta^2, \{\tau_k\}_{k=1}^K, \{\sigma_k^2\}_{k=1}^K$
**Result:** rate allocation $\{r_k^*\}_{k=1}^K$
initialization;
$r_k = 1$ for $k \in \{1, ..., K_{min}\}$ o.w. $r_k = 0$,
$\mathcal{S} = \{1, K_{min} + 1\}, \quad \mathcal{D}_0' = \mathcal{D}_0^{-1} - \sigma_\theta^{-2}$,
$\mathbf{d} = \sum_{k=1}^{K_{min}} \alpha_k(r_k)$
**while** $\mathbf{d} < \mathcal{D}_0'$ **do**
$\quad k^* = \underset{k \in \mathcal{S}}{\operatorname{argmax}}(\alpha_k(r_k + 1) - \alpha_k(r_k))$
$\quad r_{k^*} = r_{k^*} + 1$
$\quad \mathbf{d} = \mathbf{d} - \alpha_{k^*}(r_{k^*} - 1) + \alpha_{k^*}(r_{k^*})$
$\quad \mathcal{S} = \{k| r_k < r_{k-1}\} \cup \{1\}$
**end**
**Algorithm:** greedy method for rate allocation in (**P2**)

### B. Integer Relaxation Method

Let (**P'2**) be the corresponding relaxed problem of (**P2**). Solving the first order KKT necessary optimality conditions for (**P'2**) yields a similar solution to (13) as the following:

$$r_k^{\ddagger} = 0.5\left[\log_2\left(\tau_k^2(\lambda^{\ddagger} - \sigma_k^2 - \sqrt{\lambda^{\ddagger 2} - 2\lambda^{\ddagger}\sigma_k^2})^{-1}\right) - \log_2 3\right]^+ \quad (21)$$

One can show that the objective function, i.e., $\sum_{k=1}^K r_k$ is a strictly increasing function of $\alpha_k$'s. Hence, the optimal solution to (**P'2**) must satisfy the MSE constraint as equality, i.e., $\sum_{k=1}^K \alpha_k = \mathcal{D}_0'$. Using the MSE equality constraint we find that $\lambda^{\ddagger}$ in (21) is the solution to the following equation:

$$f(\lambda, \{\sigma_k^2\}_{k=1}^{K^{\ddagger}}) = \sum_{k=1}^{K^{\ddagger}} \frac{1}{\lambda - \sqrt{\lambda^2 - 2\lambda\sigma_k^2}} = \mathcal{D}_0', \quad (22)$$

where $K^{\ddagger} = \max\{k| k \geq K_{min}, \lambda^{\ddagger} > 2\sigma_{K^{\ddagger}}^2, r_k^{\ddagger} > 0\}$. Similar to what we did for the solution in (13), one can verify that the solution in (21) satisfies the second order sufficient optimality conditions in [42, p.301, proposition 3.2.1]. Note that (22) does not necessarily have a *real* solution for $\lambda$. We first let $K^{\ddagger} = K$, i.e., the largest possible value for $K^{\ddagger}$ in the feasible set $\mathcal{F} = \{K_{min}, ..., K\}$ and solve (22). If there is no *real* solution for $\lambda$ we decrease the number of active sensors by one, i.e., $K^{\ddagger} = K - 1$, and solve (22). We continue reducing the number of active sensors one by one until we reach a *real* solution for $\lambda^{\ddagger}$ or $K^{\ddagger} = K_{min}$ (the smallest possible value for $K^{\ddagger}$ in the feasible set). Even when $K^{\ddagger} = K_{min}$ it is still possible that solving (22) does not yield a *real solution* for $\lambda^{\ddagger}$. Since $f(\lambda, \{\sigma_k^2\}_{k=1}^{K_{min}})$ is an increasing function of $\lambda$, this scenario would occur when $\mathcal{D}_0' < f(\lambda, \{\sigma_k^2\}_{k=1}^{K_{min}})|_{\lambda = 2\sigma_{K_{min}}^2}$. In this scenario we let $\lambda^{\ddagger} = 2\sigma_{K_{min}}^2$. Substituting $\lambda^{\ddagger} = 2\sigma_{K_{min}}^2$ in (21) and then $r_k^{\ddagger}$'s in $\sum_{k=1}^{K_{min}} \alpha_k$, we obtain:

$$\sum_{k=1}^{K_{min}} \alpha_k = \left(\sum_{k=1}^{K_{min}} \frac{1}{\lambda - \sqrt{\lambda^2 - 2\lambda\sigma_k^2}}\right)|_{\lambda = 2\sigma_{K_{min}}^2} > \mathcal{D}_0',$$

implying that the MSE constraint is met. Using similar ap-proximation that led us to (15), we can approximate (21) as:

$$r_k^\ddagger \approx [\log_2(\frac{\tau_k}{\sigma_k^2}) + 0.5\log_2(K^\ddagger 3^{-1}(\sum_{k=1}^{K} \sigma_k^{-2} - \mathcal{D}_0')^{-1})]^+. \quad (23)$$

Equation (23) shows as target MSE approaches its feasible minimum, i.e., as $\mathcal{D}_0 \to \frac{1}{\sigma_\theta^{-2} + \sum_{k=1}^{K}\sigma_k^{-2}}$ and $\mathcal{D}_0' \to \sum_{k=1}^{K}\sigma_k^{-2}$, the rates $r_k^\ddagger$'s become very large, i.e., $r_k^\ddagger \to \infty$.

### C. Individual Rate Allocation Method

Following a similar reasoning to the one provided in Section IV-D for a homogeneous network and recalling the discussion on satisfying the MSE constraint as equality in Section V-B, we conclude that, given $\mathcal{D}_0'$, there exists an optimal number of active sensors $K^{opt}$, associated with an optimal quantization rate $r^{opt}$, where $K^{opt}\alpha(r^{opt}) = \mathcal{D}_0'$ and our problem is to minimize $K^{opt}r^{opt}$ subject to this MSE equality constraint. This optimization problem for a heterogeneous network, reduces to almost the same as in (20), with a difference that the search domain includes any positive integer number, i.e.,

$$r_k^{sopt} = \operatorname*{argmin}_{r \in \mathbb{Z}^+}\{r(\sigma_k^2 + \tau_k^2 3^{-1}(2^r - 1)^{-2})\}, \quad (24)$$

$$K^{sopt} = \min\{k | K_{min} \leq k \leq K, \sum_{i=1}^{k}\alpha_i(r_i) \geq \mathcal{D}_0'\}. \quad (25)$$

Note that there is no need to solve (24) for all sensors, since the rate allocation continues only till we find $K^{sopt}$ in (25).

## VI. CRAMÉR RAO LOWER BOUND

We derive the CRLB for any Baysian estimator of $\theta$ based on quantized observations $\{m_k\}_{k=1}^{K}$. Assuming that the regularity condition is satisfied, i.e., $\mathbb{E}\{\frac{\partial \ln p(\mathbf{m},\theta)}{\partial \theta}\} = 0$ [36] we write the Fisher information:

$$F = -\mathbb{E}\{\frac{\partial^2 \ln p(\mathbf{m},\theta)}{\partial^2 \theta}\} = -\mathbb{E}\{\frac{\partial^2 \ln p(\mathbf{m}|\theta)}{\partial^2 \theta}\} - \mathbb{E}\{\frac{\partial^2 \ln p(\theta)}{\partial^2 \theta}\}. \quad (26)$$

Assuming that $m_k$'s conditioned on $\theta$ are independent, i.e., $\ln p(\mathbf{m}|\theta) = \sum_{k=1}^{K}\ln p(m_k|\theta)$, the first and second derivatives of the log-likelihood function become:

$$\frac{\partial \ln p(\mathbf{m}|\theta)}{\partial \theta} = \sum_{k=1}^{K}\frac{1}{p(m_k|\theta)}\frac{\partial p(m_k|\theta)}{\partial \theta},$$

$$\frac{\partial^2 \ln p(\mathbf{m}|\theta)}{\partial^2 \theta} = \underbrace{\sum_{k=1}^{K}\frac{1}{p(m_k|\theta)}\frac{\partial^2 p(m_k|\theta)}{\partial^2 \theta}}_{=F_a}$$
$$-\underbrace{\sum_{k=1}^{K}\frac{1}{p^2(m_k|\theta)}(\frac{\partial p(m_k|\theta)}{\partial \theta})^2}_{=F_b}.$$

In the following, we find $\mathbb{E}\{F_a\}, \mathbb{E}\{F_b\}$. We have:

$$\mathbb{E}\{F_a\} = \sum_{k=1}^{K}\int p(\theta)\frac{\partial^2(\overbrace{\sum_{i=1}^{M_k}s_{k,i}(\theta)}^{=1})}{\partial^2 \theta}d\theta = 0,$$

$$\mathbb{E}\{F_b\} = \sum_{k=1}^{K}\int p(\theta)\sum_{i=1}^{M_k}\frac{1}{s_{k,i}(\theta)}(\dot{s}_{k,i}(\theta))^2 d\theta,$$

where $s_{k,i}(\theta) = p(m_k = m_{k,i}|\theta) = \mathrm{p}\{m_{k,i} - \frac{\Delta_k}{2} \leq h_k\theta + n_k \leq m_{k,i} + \frac{\Delta_k}{2}|\theta\}$ and $\dot{s}_{k,i}(\theta) = \frac{\partial s_{k,i}(\theta)}{\partial \theta}$. To complete the derivations of $F$ we need to characterize $s_{k,i}(\theta)$ and $\dot{s}_{k,i}(\theta)$. Combining all above and recalling $\theta \sim \mathcal{N}(0,\sigma_\theta^2)$, we obtain:

$$F = \frac{1}{\sigma_\theta}\sum_{k=1}^{K}\sum_{i=1}^{M_k}\int \frac{(\dot{s}_{k,i}(\theta))^2}{s_{k,i}(\theta)}\phi(\frac{\theta}{\sigma_\theta})d\theta + \frac{1}{\sigma_\theta^2}, \quad (27)$$

where $\phi(.)$ is the standard normal probability density function (pdf). Equation (27) is true for arbitrarily distributed $h_k$'s with $\mathbb{E}\{h_k\} = 1 \ \forall k$, and $\mathrm{var}(h_k) = \sigma_{h_k}^2$. When $h_k$'s are Gaussian we have:

$$s_{k,i}^G(\theta) = \Phi(\frac{\zeta_{k,i+1} - \theta}{\sqrt{\theta^2\sigma_{h_k}^2 + \sigma_{n_k}^2}}) - \Phi(\frac{\zeta_{k,i} - \theta}{\sqrt{\theta^2\sigma_{h_k}^2 + \sigma_{n_k}^2}}),$$

where $\zeta_{k,i} = m_{k,i} - \frac{\Delta_k}{2}$, $\zeta_{k,i+1} = m_{k,i} + \frac{\Delta_k}{2}$ are the quantizer boundaries, and $\Phi(.)$ is the cumulative distribution function (CDF) of a standard normal random variable. Deriving $\dot{s}_{k,i}^G(\theta)$ is straightforward and reduces to subtraction of two scaled standard normal pdf.

## VII. EXTENSION TO ERRONEOUS CHANNELS

To obtain our results so far we have focused on error-free communication channel model, i.e., the quantization bits from the sensors are available at the FC, to feed the LMMSE estimator. The results can be extended to independent BSCs with different error probabilities $p_k$. Suppose sensor $k$ uses binary natural coding (BNC) to code its quantized message $m_k$, that is sent through a BSC with error probability $p_k$, and $\hat{m}_k$ is the corresponding recovered quantization level at the FC, where in general $\hat{m}_k \neq m_k$, due to channel errors.

### A. LMMSE Estimator and its corresponding MSE

The LMMSE estimator and its MSE would have the same forms as in (4) and (5), with the difference that vector $\boldsymbol{m}$ is replaced with vector $\hat{\boldsymbol{m}}$. We characterize $\mathbb{E}\{\theta\hat{m}_k\}$ and $\mathbb{E}\{\hat{m}_k\hat{m}_l\}$ as the following:

$$\mathbb{E}\{\theta\hat{m}_k\} = \mathbb{E}\{\mathbb{E}\{\theta\hat{m}_k|\theta,m_k\}\} = \mathbb{E}\{\theta\mathbb{E}\{\hat{m}_k|m_k\}\}. \quad (28)$$

With BNC of bit sequences and BSC model we have $\mathbb{E}\{\hat{m}_k|m_k\} = (1 - 2p_k)m_k$ [45]. Thus (28) reduces to $\mathbb{E}\{\theta\hat{m}_k\} = (1-2p_k)\mathbb{E}\{\theta m_k\}$, where $\mathbb{E}\{\theta m_k\}$ is characterized in (6). For $\mathbb{E}\{\hat{m}_k\hat{m}_l\}$, $k \neq l$ and $k = l$ we have:

$$\mathbb{E}\{\hat{m}_k\hat{m}_l\} = \mathbb{E}\{\mathbb{E}\{\hat{m}_k\hat{m}_l|\theta,m_k,m_l\}\} \stackrel{(a)}{=}$$
$$\mathbb{E}\{\mathbb{E}\{\hat{m}_k|m_k\}\mathbb{E}\{\hat{m}_l|m_l\}\} = (1-2p_k)(1-2p_l)\mathbb{E}\{m_km_l\},$$
$$\mathbb{E}\{\hat{m}_k^2\} = \mathbb{E}\{\mathbb{E}\{\hat{m}_k^2|m_k\}\} \stackrel{(b)}{=} g_k\mathbb{E}\{m_k^2\} + R_k, \quad (29)$$

where (a) in (29) is obtained using the facts that $(i)$ given $m_k, m_l$ then $\hat{m}_k, \hat{m}_l$ are independent, $(ii)$ given $\theta$, then $m_k, m_l$ are uncorrelated (since $n_k, n_l, h_k, h_l$ are all uncorrelated). And $g_k = (1-p_k)^{r_k-1}(1+p_k(r_k-5))$ and $R_k = (4/3)(1-p_k)^{r_k-1} p_k \tau_k^2 (2^{r_k}+1)(2^{r_k}-1)^{-1}$. To obtain (b) in (29) we assume at most one bit in a sequence of $r_k$ bits can be flipped due to the channel errors (roughly speaking $p_k \ll r_k^{-1}$). This is a reasonable assumption noting that for a poor channel with $p_k \approx 0.1$ and typical quantization rates of $r_k \leq 6$, flipping more than one bit in an $r_k$-bit sequence is unlikely [45]. Note that $\mathbb{E}\{m_k m_l\}$ and $\mathbb{E}\{m_k^2\}$ in (29) are characterized in (7). Having (28), (29), the LMMSE estimator and its corresponding MSE are characterized for BSC model.

### B. CRLB and Fisher Information Expressions

To find $F$ based on $\hat{m}_k$'s, we need to find the counterpart of (26), where $\boldsymbol{m}$ is replaced with $\hat{\boldsymbol{m}}$. For independent BSC model, $\hat{m}_k$'s conditioned on $\theta$ would be independent, leading to $\ln p(\hat{\boldsymbol{m}}|\theta) = \sum_{k=1}^{K} \ln p(\hat{m}_k|\theta)$. Following similar steps as in Section VI, we find new $F_a$ to be zero. New $F_b$ can be found by replacing $p(m_k = m_{k,i}|\theta)$ with $p(\hat{m}_k = m_{k,i}|\theta)$ in the derivations. All that remains is to characterize:

$$p(\hat{m}_k = m_{k,i}|\theta) = \sum_{j=1}^{M_k} e_k^{ji} \, p(m_k = m_{k,j}|\theta),$$

where $e_k^{ij}$ is the probability of receiving level $m_{k,i}$, while level $m_{k,j}$ is transmitted from sensor $k$. Note that $e_k^{ij}$ can be found in terms of $p_k$, i.e., $e_k^{ji} = (p_k)^{n(j,i,r_k)}(1-p_k)^{r_k - n(j,i,r_k)}$, where $n(j,i,r_k)$ is the Hamming distance between BNC representations of $m_{k,j} = \sum_{l=1}^{r_k} b_{k,j,l} 2^{r_k - l}$ and $m_{k,i} = \sum_{l=1}^{r_k} b_{k,i,l} 2^{r_k - l}$. To sum up, $F$ becomes:

$$F = \sigma_\theta^{-1} \sum_{k=1}^{K} \sum_{i=1}^{M_k} \int \frac{\left(\sum_{j=1}^{M_k} e_k^{ji} \, \dot{s}_{k,j}(\theta)\right)^2}{\left(\sum_{j=1}^{M_k} e_k^{ji} \, s_{k,j}(\theta)\right)} \, \phi\!\left(\frac{\theta}{\sigma_\theta}\right) d\theta + \sigma_\theta^{-2}.$$

## VIII. NUMERICAL AND SIMULATION RESULTS

In this section, we corroborate our analytical results with numerical simulations. These results validate the accuracy of our analysis and illustrate the effectiveness and superiority of the proposed rate allocation schemes. We consider networks of sizes $K = 5, 10, 50$ and conduct simulations for over $10^5$ observation channels with randomly generated $\{\sigma_{n_k}^2, \sigma_{h_k}^2\}_{k=1}^{K}$ and depict the average performance for all rate allocation methods (greedy, integer relaxation (relaxed), individual rate allocation (IRA), order aware (OA) uniform, and uniform). We generate $\sigma_{n_k}^2$ such that $\mathbb{E}\{\sigma_{n_k}^2\} = 1$ or $\mathbb{E}\{\sigma_{n_k}^2\} = k_n$. To investigate the effect of multiplicative observation noise variance on the network dynamics and performance, we let $\mathbb{E}\{\sigma_{h_k}^2\} = k_h = 0.1, 1, 2, 4$ to indicate low, moderate, high, and very high multiplicative noise variance.

Figs. 2a and 2b compare the analytical MSE in (9) and simulated MSE for $K = 10, 50$, when greedy method is employed[4]. The simulations are conducted for $h_k$'s drawn from

[4]Integer relaxation and IRA exhibit similar results, and the plots are omitted, for the sake of saving space.

Gaussian, uniform, and Laplacian distributions. We observe that the analytical MSE is a good approximation of simulated MSE for almost all scenarios, and the approximation accuracy improves as $K$ increases and/or $k_h$ decreases. Also, except for small $K$ and very high $k_h$, the distribution of $h_k$ has negligible effect on the approximation accuracy.

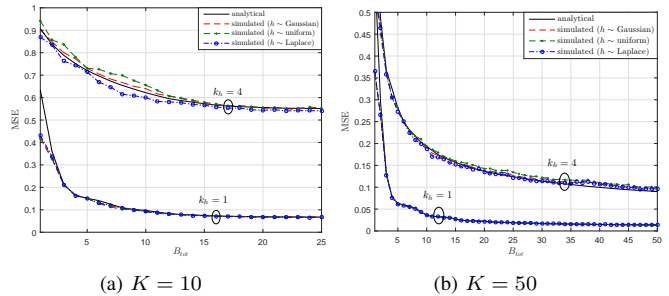

(a) $K = 10$  (b) $K = 50$

Fig. 2: Analytical and simulated MSE of greedy vs. $B_{tot}$

Figs. 3 compares the analytical MSE for different methods, $K = 5$ and $k_h = 0.1, 1, 2$. We observe the MSE performance gap between uniform (including OA uniform), and greedy and integer relaxation are remarkable. As $B_{tot}$ increases, the performance of greedy, integer relaxation, uniform, OA uniform approaches that of the clairvoyant centralized estimation. However, there is a persistent gap between the performance of IRA and the clairvoyant case, even for large $B_{tot}$. The performance of greedy and integer relaxation are almost the same for all scenarios. Similar observations are valid for $K = 50$ and the plots are omitted due to lack of space. For large $K$ and small $k_h$ the performance of the individual rate allocation competes with greedy and integer relaxation methods, however, for small $K$ or high $k_h$ it loses the competition. On the other hand, when $B_{tot}$ is relatively small compared to $K$, greedy, integer relaxation, and IRA have the same performance. As expected, we observe larger $k_h$ (larger $K$) leads to a larger (smaller) MSE for all methods.

Fig. 4 depicts the MSE performance of different methods and the associated CRLB for $K = 5$, where $h_k$'s are drawn from Gaussian distribution for CRLB. The $\sigma_{n_k}^2$'s and $\sigma_{h_k}^2$'s are independently generated with Chi-Square distribution $\sigma_{h_k}^2 \sim \chi^2(k_h)$, $\sigma_{n_k}^2 \sim \chi^2(1)$. For $k_h = 1$ (moderate multiplicative noise), there is a noticeable gap between the MSE and the associated CRLB for all methods, whereas for $k_h = 0.1$ (low noise) and large $B_{tot}$, this gap tends to be very small. This is in agreement with the result that the MSE of MMSE estimator for a Gaussian linear observation model achieves the CRLB [36]. In fact, for $k_h = 0$ the observation model in (1) becomes the linear Gaussian model $x_k = \theta + n_k$ and when $B_{tot} \to \infty$, LMMSE estimator in (4) becomes MMSE estimator, which achieves the CRLB. Similar observations are valid for $K = 50$ and the plots are omitted due to lack of space.

Figs. 5a and 5b depict the number of active sensors versus $B_{tot}$ for all methods and $K = 10, 50$. For $k_h = 2$ (high noise) more sensors become active to reduce the noise effect, by averaging over observations coming from more sensors, leading to smaller quantization rates (coarser quantization). On the other hand, for $h_k = 0.1$ (low noise) less sensors become active, leading to larger quantization rates (finer quantization). These observations illustrate the trade-off mentioned in explanations following (15) and in subsection IV-D. Note that greedy and integer relaxation methods activate fewer sensors, compared with those of IRA and uniform methods, and still provide better MSE performance (see also Fig. 3). Figs. 6a and 6b illustrate the average quantization rates of active sensors versus $B_{tot}$ for all methods and $K = 10, 50$. For $k_h = 2$ (high noise) the average quantization rates are smaller (more active sensors with coarser quantization). On the other hand, for $h_k = 0.1$ (low noise) the average quantization rates are larger (less active sensors with finer quantization). These observations illustrate the trade-off mentioned in explanations following (15) and in subsection IV-D.

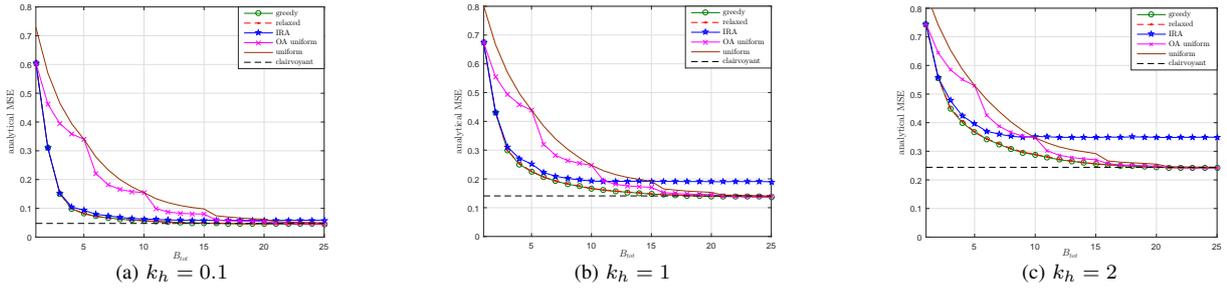

Fig. 3: MSE of different rate allocation methods vs. $B_{tot}$ for $K = 5$

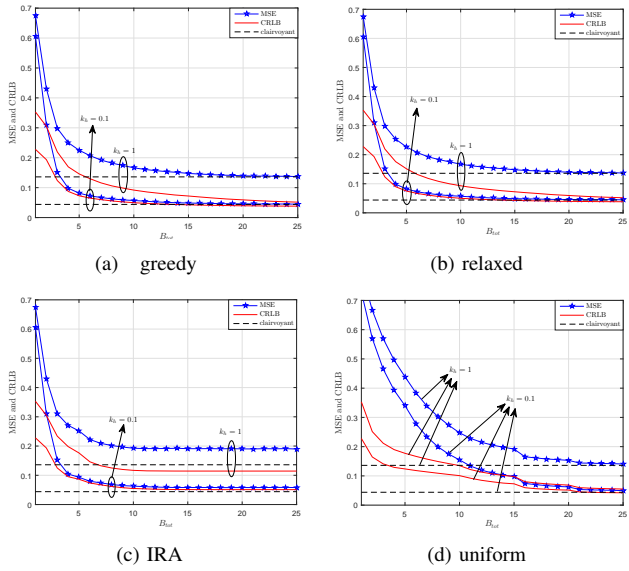

Fig. 4: MSE and CRLB vs. $B_{tot}$ for $K = 5$

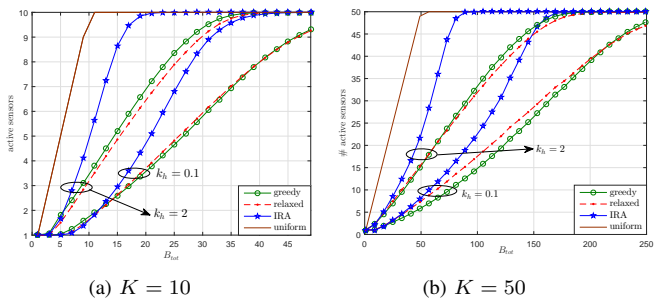

Fig. 5: Number of active sensors vs. $B_{tot}$

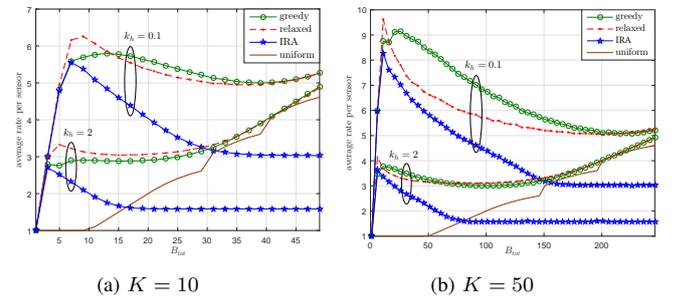

Fig. 6: Average quantization rate of active sensors vs. $B_{tot}$

Fig. 7 illustrates the required bandwidth, i.e., sum of quantization rates $\sum_{k=1}^{K} r_k$ versus a target MSE (to be satisfied), for all methods and $K = 50$. Note that greedy and integer relaxation methods require much less bandwidth to achieve the target MSE, compared with those of IRA, uniform, and OA uniform. Similar observations are valid for $K = 5$ and the plot is omitted due to lack of space. These figures (excluding IRA) show that, more bandwidth is required to (i) satisfy a smaller target MSE, (ii) satisfy a fixed target MSE for larger $k_h$, (iii) satisfy a fixed target MSE for smaller $K$. In some sub-figures the required bandwidth for some target MSE values are left blank, since the target MSE is not achievable for that particular network setting. Note that IRA method is different from greedy and integer relaxation methods, since it is blind to target MSE and $K$ value (see (24),(25)), i.e., the assigned quantization rates are independent of the target MSE and $K$ value and the number of active sensors is kept at minimum, such that the target MSE is satisfied. For illustrative purposes, consider a large and easy-to-be-satisfied MSE target, such that it lies in the interval $[0.7, 1]$. Such a target MSE most likely can be satisfied with one active sensor (see Figs. 8a, 8b), and one bit (see Fig. 9) in greedy and integer relaxation methods. However, since IRA is blind to the target MSE, it assigns a quantization rate to the only active sensor, according to the observation channel quality $\sigma_1^2$, that is likely to be larger than one bit (in fact, the smaller $k_h$ is, the larger $r_k$ is).

Figs. 8a and 8b depict the number of active sensors when the target MSE is met for all methods and $K = 10, 50$. Note that greedy and integer relaxation methods activate fewer sensors to satisfy the target MSE, compared with those of IRA and uniform and OA uniform. For $k_h = 2$ (high noise), all methods require more active sensors to satisfy the target MSE (similar observations to those of Fig.5). Figs. 9a and 9b illustrate the average quantization rates of active sensors versus target MSE, for all methods and $K = 10, 50$. Similar conclusions to those for Fig. 7 can be made here. Note that IRA loses the competition to uniform methods for large target MSE values. These figures show that, larger average quantization rate is required to $(i)$ satisfy a smaller target MSE, $(ii)$ satisfy a fixed target MSE for larger $k_h$, $(iii)$ satisfy a fixed target MSE for smaller $K$ (compare the average rate of all algorithms except IRA in Fig. 9). In the figures the average quantization rates for some target MSE values are left blank, since the target MSE is not attainable for that particular network setting. Combining the observations from Figs. 7, 8, 9, we conclude that IRA method is not suitable to address (**P2**). To show the effect of erroneous communication channels, Fig. 10 depicts the analytical and simulated MSE and compare them with CRLB for $p = 10^{-1}, 10^{-2}, 10^{-4}$, $K = 50$, and $k_h = 1$, when greedy method is employed. As expected, the analytical MSE is very accurate unless for large error probability $p = 0.1$ (this is expected since to derive (29), we assume $p_k \ll r_k^{-1}$, which is not true for $p = 0.1$).

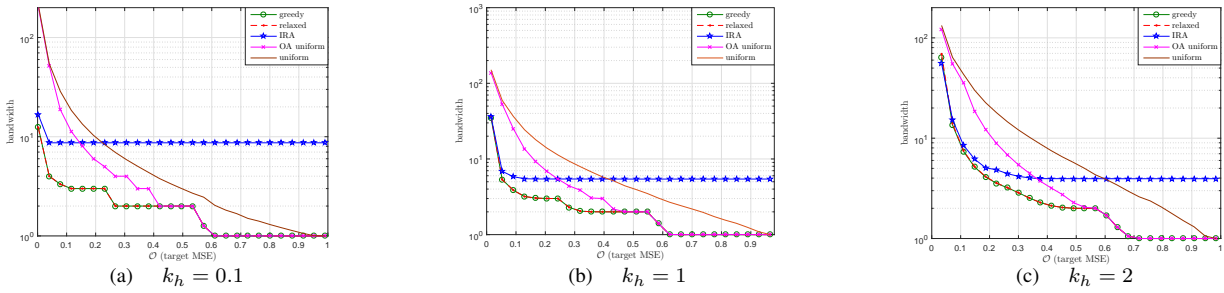

Fig. 7: Required bandwidth of different rate allocation methods vs. target MSE for $K = 50$

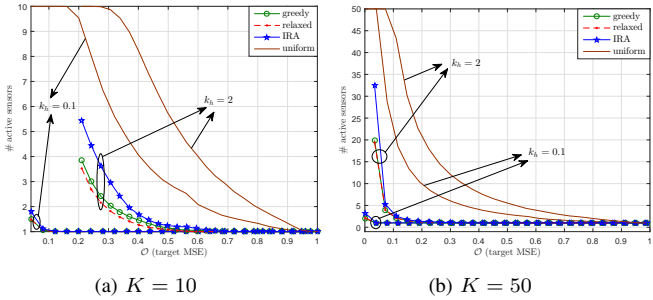

Fig. 8: Number of active sensors vs. target MSE

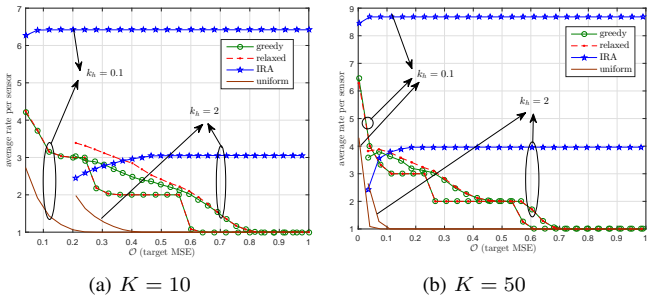

Fig. 9: Quantization rate of active sensors vs. target MSE

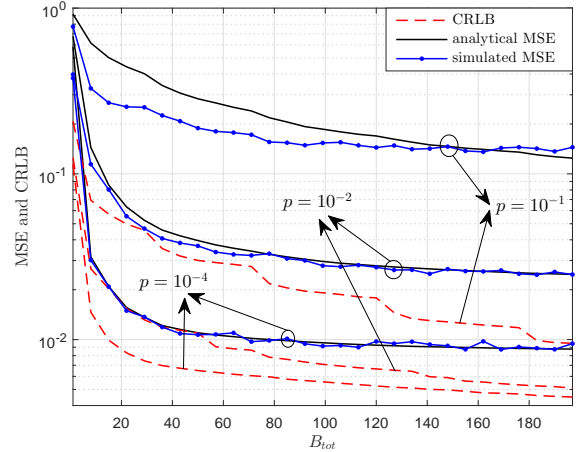

Fig. 10: MSE and CRLB of greedy vs. $B_{tot}$ for $K = 50$

• *Comparison with Rate-Distortion (R−D) Bound in [27]*: As we mentioned in Section I, DES and the quadratic Gaussian CEO problem are different and hence the $R$–$D$ bounds in [27]–[29] are less relevant to the problem in hand. Interestingly, our simulations show that in some scenarios, even the CRLB based on multi-bit quantization can reach the $R - D$ bound in [27]. Fig. 15 compares the CRLB based on one-bit and multi–bit quantization with the $R-D$ bound for different $K$ and $k_n$. We use the $R-D$ bound in [27] given below, which is for a heterogeneous network with limited number of agents:

$$R(D) = 0.5 \left[ \log_2\left(\frac{\sigma_\theta^2}{D} \prod_{k=1}^{\tilde{K}} \left( \frac{\tilde{K}}{\sigma_{n_k}^2 \left(\frac{1}{\tilde{D}(\tilde{K})} - \frac{1}{D}\right)} \right) \right) \right]^+, \quad (30)$$

where $\tilde{D}(\tilde{K}) = (\sigma_\theta^{-2} + \sigma_{n_1}^{-2} + ... + \sigma_{n_{\tilde{K}}}^{-2})^{-1}$ and $\tilde{K}$ is the largest value that satisfies $\frac{\tilde{K}}{\sigma_{n_{\tilde{K}}}^2} - \left(\frac{1}{\tilde{D}(\tilde{K})} - \frac{1}{D}\right) \geq 0$. Note that the gap between the $R-D$ bound and the CRLB based on multi-bit quantization is not persistent: as $B_{tot}$ increases for a fixed $K$, the gap fades away and the latter approaches the former.

• *One-bit MLE and MAP vs. Multi-bit LMMSE:* As we mentioned in Section I, there is a significant gap between the CRLB performance based on one-bit quantization and the clairvoyant benchmark (unquantized observations are available at the FC), when the dynamic range of $\theta$ is large with respect to $\sigma_{n_k}^2$ [7]. To illustrate this, Fig. 11 plots classical CRLB $= (\frac{(\dot{s}_{k,i}(\theta))^2}{s_{k,i}(\theta)})^{-1}$ versus $\theta$ in a homogeneous network with $\sigma_n^2 = 1$, assuming $\theta \in [-4, 4]$. corresponding to one bit [34] and two bit quantization. We observe the classical CRLB corresponding to two bit quantization is significantly better than that of one bit quantization, when $\theta$ is larger than $\sigma_n^2$. One expects similar observation holds when we compare MSE of MLE corresponding to one bit and two bit quantization. In Appendix X we provide an analytical reason for the poor performance of CLRB and MLE based on one bit quantization. When pdf of $\theta$ is known *a priori*, Maximum A Posteriori estimator (MAP) can be used instead of MLE. Figs. 13 and 14 plot MSE of one-bit MLE, one-bit MAP, and the proposed multi-bit LMMSE estimators, as well as the associated (one-bit and multi-bit) CRLBs and $R-D$ for a heterogeneous network of size $K = 50$ versus $k_n$ and $B_{tot}$, respectively. As expected, in all cases the proposed multi-bit LMMSE outperforms one-bit MLE and one-bit MAP. Fig. 13 also shows that for small $\sigma_n^2$ one-bit MLE and one-bit MAP perform poorly. Also, as $\sigma_\theta^2$ becomes larger the performance gap between one-bit MLE, one-bit MAP and the proposed multi-bit LMMSE increases significantly, as expected from our analysis in Section X.

## IX. CONCLUSIONS

We considered DES of a Gaussian source in a heterogenous bandwidth constrained WSN, where the source is corrupted by independent multiplicative and additive observation noises, with incomplete statistical knowledge of the multiplicative noise. For uniform multi-bit quantizers, we derived the closed-form MSE expression for the LMMSE estimator at the FC, and verified the accuracy of our derivations via simulations. For both error-free and erroneous communication channels (using BSC model) we proposed several rate allocation methods to ($i$) minimize the MSE given a network bandwidth constraint, and ($ii$) minimize the required network bandwidth given a target MSE. We also derived the Bayesian CRLB and compared the MSE performance of our proposed methods against the CRLB. Our results corroborate that, for low power multiplicative observation noises and adequate network bandwidth, the gaps between the MSE of our proposed methods and the CRLB are negligible, while the performance of other methods like *individual rate allocation in [6]*, and *uniform* is not satisfactory. Through analysis and simulations, we showed that one-bit MLE and one-bit MAP in the literature have poor performance, when the realizations of unknown is large (compared with the observation noise variances), whereas our proposed multi-bit LMMSE significantly outperforms estimators based on one-bit quantization.

## X. APPENDIX

For a homogeneous network, the likelihood function corresponding to one bit quantization is $L(\theta) = \sum_{k=1}^{K} \ln \Phi(y_k \frac{\theta}{v(\theta)})$, where $v(\theta) = \sqrt{\theta^2 \sigma_h^2 + \sigma_n^2}$ and $y_k = \text{sign}(h_k \theta + n_k)$. Therefore, the one-bit MLE is $\hat{\theta}_{ML} = \underset{\theta \in [-\theta_{max}, \theta_{max}]}{\text{argmax}} L(\theta)$. We have:

$$\dot{L}(\theta) = \frac{\partial L}{\partial \theta} = \overbrace{\frac{\sigma_n^2}{v^2(\theta)} \phi(\frac{\theta}{v(\theta)})}^{=l_a(\theta)} \overbrace{\left[\frac{N^+}{\Phi(\frac{\theta}{v(\theta)})} - \frac{N^-}{\Phi(\frac{-\theta}{v(\theta)})}\right]}^{=l_b(\theta)} \quad (31)$$

where $N^+$ and $N^-$ are number of +1's and −1's in all $y_k$'s, respectively, and $N^+ + N^- = K$. Since $l_a(\theta) > 0$, the solution of $l_b(\theta) = 0$ is $\hat{\theta}_{ML}$. However, in some cases this equation has no solution in $[-\theta_{max}, \theta_{max}]$ and $L(\theta)$ is strictly monotonic. One instance is when both multiplicative and additive noises are very small, such that all $y_k$'s are equal to +1 almost surely, i.e., $N^+ = K, N^- = 0$, and hence $\hat{\theta}_{ML} = \theta_{max}$. This poor performance of MLE for low power noises is unintuitive. Below, we provide an analytical explanation for this. Let $P_s$ denote the probability of $L(\theta)$ being strictly monotonic (i.e., the probability of $l_b(\theta) > 0 \lor l_b(\theta) < 0$). We find:

$$\begin{aligned} P_s &= \left[\Phi^{K-q}(\frac{\theta}{v(\theta)}) + \Phi^{K-q}(\frac{-\theta}{v(\theta)})\right] \quad (32) \\ &\times \sum_{i=0}^{q} \left(\Phi^i(\frac{\theta}{v(\theta)}) \Phi^{q-i}(\frac{-\theta}{v(\theta)})\right) \end{aligned}$$

where $q = \left\lfloor \Phi(\frac{-\theta_{max}}{v(\theta_{max})}) \right\rfloor$. Interestingly, $P_s$ is an even function of $\frac{|\theta|}{\sigma_n}$ and can get very close to one (plot of $P_s$ versus $\frac{|\theta|}{\sigma_n}$ is deleted due to lack of space). In fact, when $\frac{|\theta|}{\sigma_n}$ is large enough MLE becomes independent of $y_k$'s and realization of $\theta$, i.e., $\hat{\theta}_{ML} = \pm \theta_{max}$, leading to poor estimation performance. The larger is the dynamic range of $\theta$ (compared with $\sigma_n^2$), the larger is the estimation error of one-bit MLE. When pdf of $\theta$ is known *a priori*, MAP can be used instead of MLE, where $\hat{\theta}_{MAP} = \text{argmax}[L(\theta) + \ln(\sigma_\theta^{-1} \phi(\frac{\theta}{\sigma_\theta}))]$, i.e., $\hat{\theta}_{MAP}$ is the solution of $\dot{L}(\theta) = \frac{\theta}{\sigma_\theta^2}$. As we discussed before, in some cases $\dot{L}(\theta)$ becomes almost independent of $y_k$'s and realization of $\theta$, leading to poor estimation performance. Figs. 12.a and 12.b depict Mont Carlo averages of $\hat{\theta}_{ML}$ and $\hat{\theta}_{MAP}$ versus realizations of $\theta$ for $\sigma_n^2 = 1, \sigma_h^2 = 0.1, 1, K = 5, 50$. These figures show that as the realization of $\theta$ becomes large enough, the estimates become more inaccurate, i.e., $\hat{\theta}_{ML}$ approaches $\pm \theta_{max}$ and $\hat{\theta}_{MAP}$ gets independent of $\theta$ realization, and is determined[5] by $\sigma_\theta^2$, $\sigma_h^2$, $K$, causing severe estimation errors.

## REFERENCES

[1] A. Sani and A. Vosoughi, "On quantizer design for distributed estimation in bandwidth constrained networks," in *2015 IEEE 82nd Vehicular Technology Conference (VTC2015-Fall)*, Sept 2015, pp. 1–2.
[2] A. Vempaty, H. He, B. Chen, and P. Varshney, "On quantizer design for distributed bayesian estimation in sensor networks," *Signal Processing, IEEE Transactions on*, vol. 62, no. 20, pp. 5359–5369, Oct 2014.
[3] S. Kar, H. Chen, and P. Varshney, "Optimal identical binary quantizer design for distributed estimation," *Signal Processing, IEEE Transactions on*, vol. 60, no. 7, pp. 3896–3901, July 2012.
[4] J. Fang and H. Li, "Hyperplane-based vector quantization for distributed estimation in wireless sensor networks," *Information Theory, IEEE Transactions on*, vol. 55, no. 12, pp. 5682–5699, Dec 2009.

[5]In contrast to $\hat{\theta}_{ML}$, our simulations suggest, provided that $\theta_{max} > 3\sigma_\theta$, $\hat{\theta}_{MAP}$ does not depend on $\theta_{max}$ in these cases.

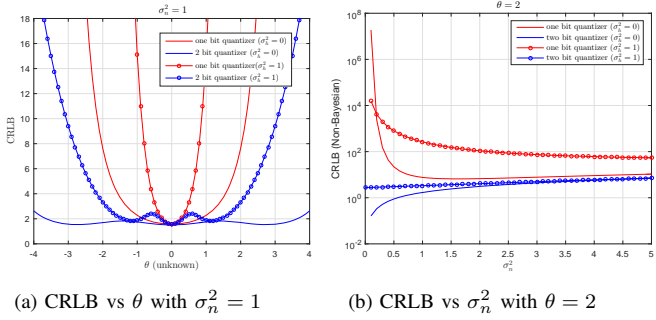

(a) CRLB vs $\theta$ with $\sigma_n^2 = 1$

(b) CRLB vs $\sigma_n^2$ with $\theta = 2$

Fig. 11: CRLB correspoding to one bit and two bit quantizers

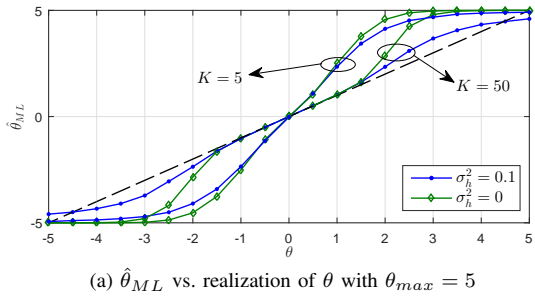

(a) $\hat{\theta}_{ML}$ vs. realization of $\theta$ with $\theta_{max} = 5$

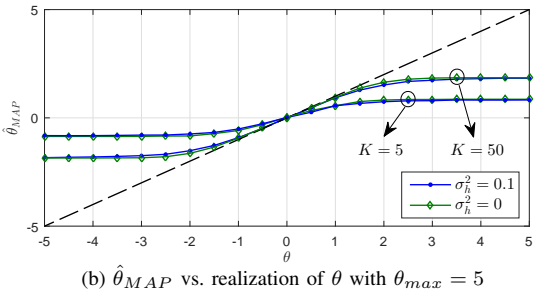

(b) $\hat{\theta}_{MAP}$ vs. realization of $\theta$ with $\theta_{max} = 5$

Fig. 12: $\hat{\theta}_{ML}$ and $\hat{\theta}_{MAP}$ vs. realizations of $\theta$

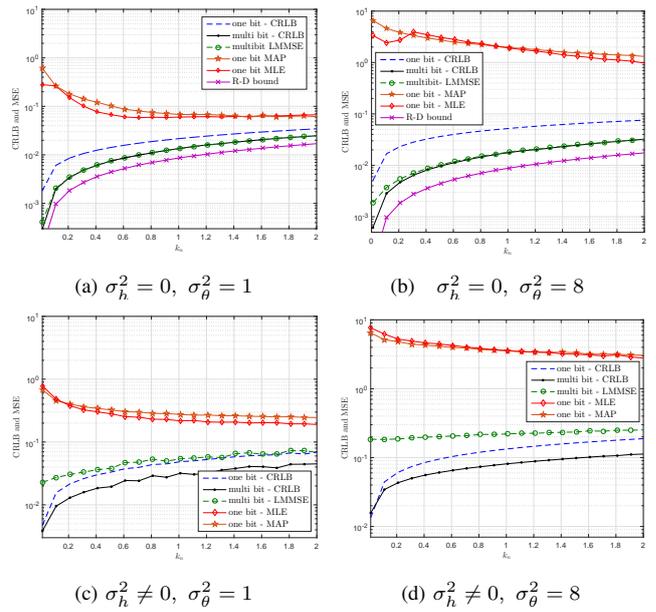

(a) $\sigma_h^2 = 0$, $\sigma_\theta^2 = 1$

(b) $\sigma_h^2 = 0$, $\sigma_\theta^2 = 8$

(c) $\sigma_h^2 \neq 0$, $\sigma_\theta^2 = 1$

(d) $\sigma_h^2 \neq 0$, $\sigma_\theta^2 = 8$

Fig. 13: Performance comparison of one-bit and multi-bit estimation vs. $k_n$, with $K = 50, B_{tot} = 50, \theta_{max} = 2\sigma_\theta$

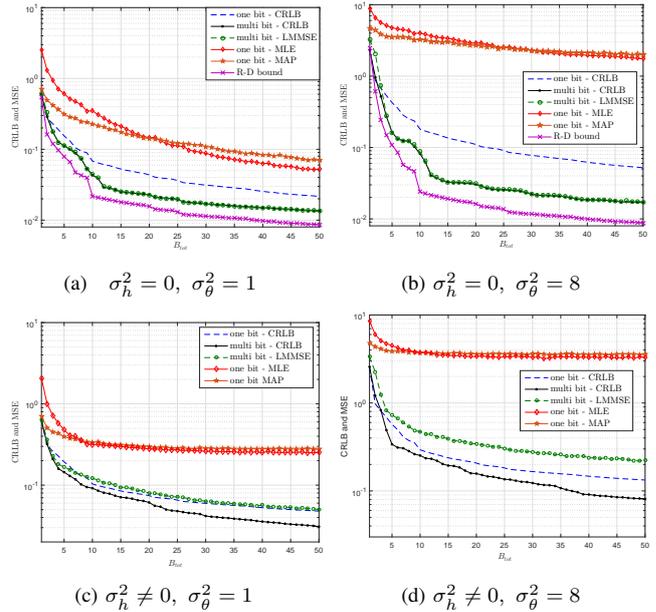

(a) $\sigma_h^2 = 0$, $\sigma_\theta^2 = 1$

(b) $\sigma_h^2 = 0$, $\sigma_\theta^2 = 8$

(c) $\sigma_h^2 \neq 0$, $\sigma_\theta^2 = 1$

(d) $\sigma_h^2 \neq 0$, $\sigma_\theta^2 = 8$

Fig. 14: Performance comparison of one bit and multi bit estimation vs. $B_{tot}$, $K = 50$, $\theta_{max} = 2\sigma_\theta$

[5] ——, "Distributed adaptive quantization for wireless sensor networks: From delta modulation to maximum likelihood," *IEEE Transactions on Signal Processing*, vol. 56, no. 10, pp. 5246–5257, Oct 2008.

[6] J. Li and G. AlRegib, "Rate-constrained distributed estimation in wireless sensor networks," *Signal Processing, IEEE Transactions on*, vol. 55, no. 5, pp. 1634–1643, May 2007.

[7] A. Ribeiro and G. Giannakis, "Bandwidth-constrained distributed estimation for wireless sensor networks-part i: Gaussian case," *Signal Processing, IEEE Transactions on*, vol. 54, no. 3, pp. 1131–1143, March 2006.

[8] S. Marano, V. Matta, and P. Willett, "Quantizer precision for distributed estimation in a large sensor network," *Signal Processing, IEEE Transactions on*, vol. 54, no. 10, pp. 4073–4078, Oct 2006.

[9] O. Dabeer and A. Karnik, "Signal parameter estimation using 1-bit dithered quantization," *IEEE Transactions on Information Theory*, vol. 52, no. 12, pp. 5389–5405, Dec 2006.

[10] Z.-Q. Luo, "Universal decentralized estimation in a bandwidth constrained sensor network," *IEEE Transactions on Information Theory*, vol. 51, no. 6, pp. 2210–2219, June 2005.

[11] A. Sani and A. Vosoughi, "On distributed vector estimation for power and bandwidth constrained wireless sensor networks," *IEEE Transactions on Signal Processing*, vol. PP, no. 99, pp. 1–1, 2016.

[12] M. Ahmed, T. Al-Naffouri, M.-S. Alouini, and G. Turkiyyah, "The effect of correlated observations on the performance of distributed estimation," *Signal Processing, IEEE Transactions on*, vol. 61, no. 24, pp. 6264–6275, Dec 2013.

[13] S. Talarico, N. Schmid, M. Alkhweldi, and M. Valenti, "Distributed estimation of a parametric field: Algorithms and performance analysis," *Signal Processing, IEEE Transactions on*, vol. 62, no. 5, pp. 1041–1053, March 2014.

[14] I. Nevat, G. Peters, and I. Collings, "Random field reconstruction with quantization in wireless sensor networks," *Signal Processing, IEEE Transactions on*, vol. 61, no. 23, pp. 6020–6033, Dec 2013.

[15] T. C. Aysal and K. E. Barner, "Constrained decentralized estimation over noisy channels for sensor networks," *IEEE Transactions on Signal Processing*, vol. 56, no. 4, pp. 1398–1410, April 2008.

[16] J.-J. Xiao, S. Cui, Z.-Q. Luo, and A. Goldsmith, "Power scheduling of universal decentralized estimation in sensor networks," *Signal Processing, IEEE Transactions on*, vol. 54, no. 2, pp. 413–422, Feb 2006.

[17] A. Behbahani, A. Eltawil, and H. Jafarkhani, "Decentralized estimation under correlated noise," *Signal Processing, IEEE Transactions on*, vol. 62, no. 21, pp. 5603–5614, Nov 2014.

[18] A. Sani and A. Vosoughi, "Resource allocation optimization for dis-

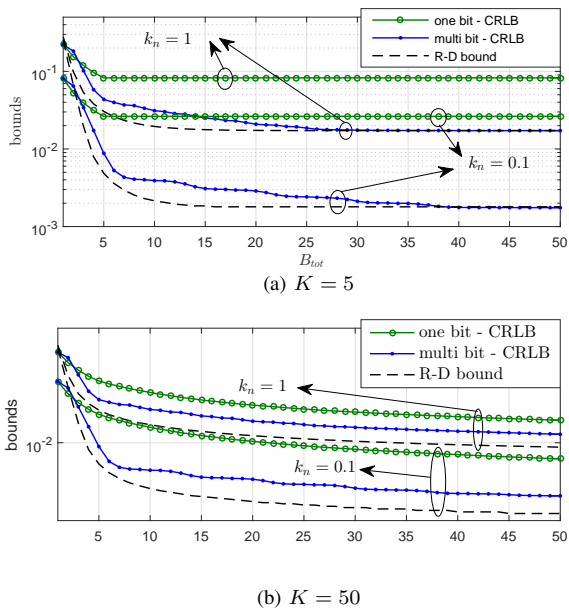

Fig. 15: $R-D$ and CRLB for $K=5, 50$ and $\sigma_\theta^2 = 1$

tributed vector estimation with digital transmission," in *Signals, Systems and Computers, 2014 48th Asilomar Conference on*, Nov 2014, pp. 1463–1467.

[19] ——, "Bandwidth and power constrained distributed vector estimation in wireless sensor networks," in *Military Communications Conference, MILCOM 2015 - 2015 IEEE*, Oct 2015, pp. 1164–1169.

[20] ——, "Noise enhanced distributed bayesian estimation," in *2017 IEEE International Conference on Acoustics, Speech and Signal Processing (ICASSP)*, March 2017, pp. 4217–4221.

[21] J. Proakis, *Digital Communications*, ser. Electrical engineering series. McGraw-Hill, 2001. [Online]. Available: https://books.google.com/books?id=sbr8QwAACAAJ

[22] K. Kotobi and S. G. Bilén, "Spectrum sharing via hybrid cognitive players evaluated by an m/d/1 queuing model," *EURASIP Journal on Wireless Communications and Networking*, vol. 2017, no. 1, p. 85, May 2017. [Online]. Available: https://doi.org/10.1186/s13638-017-0871-x

[23] S. Hatamnia, S. Vahidian, S. Assa, B. Champagne, and A. Ahmadian-Attari, "Network-coded two-way relaying in spectrum-sharing systems with quality-of-service requirements," *IEEE Transactions on Vehicular Technology*, vol. 66, no. 2, pp. 1299–1312, Feb 2017.

[24] M. K. Kiskani and B. H. Khalaj, "Novel power control algorithms for underlay cognitive radio networks," in *2011 21st International Conference on Systems Engineering*, Aug 2011, pp. 206–211.

[25] B. S. M. Mousaie, M. Soltanian, "Comsens: Exploiting pilot diversity for pervasive integration of communication and sensing in mimo-tdd-frameworks," in *MILCOM*, 2017.

[26] H. Viswanathan and T. Berger, "The quadratic gaussian ceo problem," *IEEE Transactions on Information Theory*, vol. 43, no. 5, pp. 1549–1559, Sep 1997.

[27] J. Chen, X. Zhang, T. Berger, and S. B. Wicker, "An upper bound on the sum-rate distortion function and its corresponding rate allocation schemes for the ceo problem," *IEEE Journal on Selected Areas in Communications*, vol. 22, no. 6, pp. 977–987, Aug 2004.

[28] Y. Oohama, "Rate-distortion theory for gaussian multiterminal source coding systems with several side informations at the decoder," *IEEE Transactions on Information Theory*, vol. 51, no. 7, pp. 2577–2593, July 2005.

[29] H. Behroozi and M. R. Soleymani, "Distortion sum-rate performance of successive coding strategy in quadratic gaussian ceo problem," *IEEE Transactions on Wireless Communications*, vol. 6, no. 12, pp. 4361–4365, December 2007.

[30] Y. C. Eldar, A. Ben-Tal, and A. Nemirovski, "Robust mean-squared error estimation in the presence of model uncertainties," *IEEE Transactions on Signal Processing*, vol. 53, no. 1, pp. 168–181, Jan 2005.

[31] Y. C. Eldar, "Minimax estimation of deterministic parameters in linear models with a random model matrix," *IEEE Transactions on Signal Processing*, vol. 54, no. 2, pp. 601–612, Feb 2006.

[32] A. Wiesel, Y. C. Eldar, and A. Yeredor, "Linear regression with gaussian model uncertainty: Algorithms and bounds," *IEEE Transactions on Signal Processing*, vol. 56, no. 6, pp. 2194–2205, June 2008.

[33] J. Zhu, X. Wang, X. Lin, and Y. Gu, "Maximum likelihood estimation from sign measurements with sensing matrix perturbation," *IEEE Transactions on Signal Processing*, vol. 62, no. 15, pp. 3741–3753, Aug 2014.

[34] J. Zhu, X. Lin, R. S. Blum, and Y. Gu, "Parameter estimation from quantized observations in multiplicative noise environments," *IEEE Transactions on Signal Processing*, vol. 63, no. 15, pp. 4037–4050, Aug 2015.

[35] V. Naghshin, M. C. Reed, and Y. Liu, "On the performance analysis of finite wireless network," in *2015 IEEE International Conference on Communications (ICC)*, June 2015, pp. 6530–6535.

[36] S. Kay, *Fundamentals of statistical signal processing: estimation theory*. Prentice Hall,Upper Saddle River,NJ, 1993.

[37] B. Widrow and I. Kollar, *Quantization Noise: Roundoff Error in Digital Computation, Signal Processing, Control and Communications*. Cambridge Univ. Press, 2008.

[38] R. Garnier and J. Taylor, *Discrete Mathematics: Proofs, Structures and Applications, Third Edition*, 3rd ed. Bristol, PA, USA: Taylor & Francis, Inc., 2009.

[39] G. E. Andrews, *The Theory of Partitions*. Cambridge University Press, 1984, cambridge Books Online. [Online]. Available: http://dx.doi.org/10.1017/CBO9780511608650

[40] S. Joshi and S. Boyd, "Sensor selection via convex optimization," *Signal Processing, IEEE Transactions on*, vol. 57, no. 2, pp. 451–462, Feb 2009.

[41] A. Sani, M. M. Feghhi, and A. Abbasfar, "Discrete bit loading and power allocation for ofdma downlink with minimum rate guarantee for users," *AEU - International Journal of Electronics and Communications*, vol. 68, no. 7, pp. 602 – 610, 2014. [Online]. Available: http://www.sciencedirect.com/science/article/pii/S1434841114000077

[42] D. P. Bertsekas, *Nonlinear Programming*. Belmont, MA: Athena Scientific, 1999.

[43] A. Sani and A. Abbasfar, "Qos-aware discrete bit loading for ofdma networks," in *2013 International Conference on Computing, Networking and Communications (ICNC)*, Jan 2013, pp. 994–998.

[44] ——, "Discrete bit-loading for ofdma downlink with qos consideration," in *2012 15th International Telecommunications Network Strategy and Planning Symposium (NETWORKS)*, Oct 2012, pp. 1–6.

[45] H. Leung, C. Seneviratne, and M. Xu, "A novel statistical model for distributed estimation in wireless sensor networks," *IEEE Transactions on Signal Processing*, vol. 63, no. 12, pp. 3154–3164, June 2015.